\documentclass[journal]{IEEEtran}
\IEEEoverridecommandlockouts
\usepackage[ruled,linesnumbered]{algorithm2e}
\usepackage{amsmath,amsfonts}
\usepackage{cases}
\usepackage{mathtools}
\usepackage{amssymb}
\usepackage{bm}
\usepackage{relsize}
\usepackage{array}
\usepackage{textcomp}
\usepackage{stfloats}
\usepackage{url}
\usepackage{verbatim}
\usepackage{graphicx}
\usepackage{subfigure}
\usepackage{cite}
\usepackage{color}
\usepackage{makecell}
\usepackage{etoolbox}
\usepackage{xparse}
\usepackage{float}
\usepackage{balance}
\newtheorem{proposition}{\underline{Proposition}}

\allowdisplaybreaks[4]

\usepackage{microtype}
\begin{document}

\title{Hybrid Beamforming Optimization for MIMO ISAC based on Prior Distribution Information
\thanks{This paper has been presented in part at the IEEE Global Communications Conference (Globecom),  Cape Town, South Africa, Dec. 2024 \cite{bibGlobecom}.}
\thanks{The authors are with the Department of Electrical and Electronic Engineering, The Hong Kong Polytechnic University, Hong Kong SAR, China (e-mails: yizhuo-eee.wang@connect.polyu.hk; shuowen.zhang@polyu.edu.hk).}
}
\author{\IEEEauthorblockN{Yizhuo Wang} and \IEEEauthorblockN{Shuowen Zhang},~\IEEEmembership{Senior Member,~IEEE} }

\maketitle

\begin{abstract}
This paper studies a multiple-input multiple-output (MIMO) integrated sensing and communication (ISAC) system, where a multi-antenna base station (BS) transmits dual-functional signals to communicate with a multi-antenna user and simultaneously sense the \emph{unknown} and \emph{random} location information of a point target based on the reflected echo signals and the \emph{prior distribution information} on the target's location. We study a challenging case where both the BS transmitter and the BS receiver adopt a \emph{hybrid analog-digital array} with limited radio frequency (RF) chains, and sensing can only be performed based on the analog-beamformed signals at the BS receiver. Under transceiver hybrid arrays, we characterize the sensing performance by deriving the \emph{posterior Cram\'{e}r-Rao bound (PCRB)} of the mean-squared error (MSE), which is a function of the transmit hybrid beamforming and receive analog beamforming and holds for any receiver digital signal processing and estimation. Based on this, we study the joint transmit hybrid beamforming and receive analog beamforming optimization to minimize the PCRB subject to a MIMO communication rate requirement. To draw fundamental insights, we first consider a special sensing-only system, and derive the \emph{optimal solution} to each element in the transmit/receive analog beamforming matrices that minimizes the PCRB in \emph{closed form}. An alternating optimization (AO) based algorithm is then developed which is guaranteed to converge to a Karush-Kuhn-Tucker (KKT) point. Next, we study a narrowband MIMO ISAC system and devise an efficient AO-based hybrid beamforming design algorithm by leveraging weighted minimum mean-squared error (WMMSE) and feasible point pursuit successive convex approximation (FPP-SCA) techniques. Furthermore, we derive the sensing PCRB for a  wideband MIMO orthogonal frequency-division multiplexing (OFDM) ISAC system, and jointly optimize the common transmit/receive analog beamforming for all sub-carriers and individual transmit digital beamforming for each sub-carrier. 
Numerical results validate the effectiveness of our proposed hybrid beamforming designs. Moreover, it is revealed that the number of \emph{receive} RF chains has more significant impact on the sensing performance than its transmit counterpart. Under a given budget on the total number of transmit/receive RF chains at the BS, the optimal number of transmit RF chains increases as the communication rate target increases due to the non-trivial PCRB-rate trade-off. \looseness=-1
\end{abstract}

\vspace{-1mm}

\vspace{-4mm}
\section{Introduction}\label{sec_int}
\vspace{-1mm}
The sixth-generation (6G) cellular networks will be featured by a paradigm shift from conventional single-functional communication systems to future multi-functional communication systems incorporated with sensing capabilities via the integrated sensing and communication (ISAC) technology \cite{saad2020vision}. This technology is paramount for supporting critical applications such as the industrial internet-of-things (IIoT) and the low-altitude economy.\looseness=-1

To harness the full potential of ISAC, the transmit signals need to be judiciously designed. 
The sensing performance metric adopted in existing works is either the similarity with a desired beampattern \cite{liu2020joint,hua2023optimal}, which is tractable but does not explicitly reflect sensing error, or the Cram\'{e}r-Rao bound (CRB), which provides an explicit MSE lower bound \cite{liu2022cramer,hua2024mimo,song2024cramer,tang2025dual}.
However, a fundamental limitation of the CRB is that its calculation requires the \emph{exact values} of the parameters to be sensed. In practical sensing scenarios, such exact values are precisely the quantities one aims to estimate and are therefore \emph{unknown a priori}.\looseness=-1

To address this issue, we consider a more realistic scenario where \emph{prior statistical distributions} of the parameters can be obtained from historic data or inherent target properties. Consequently, we adopt the \emph{posterior Cram\'{e}r-Rao bound (PCRB)} or Bayesian Cram\'{e}r-Rao bound (BCRB) as a more practical metric \cite{van1968detection}. The PCRB depends only on the parameters' probability density function (PDF) instead of their unknown exact values and quantifies a lower bound of the MSE when prior distribution information is available for exploitation.
 With PCRB as the sensing performance metric, \cite{xu2023radar,xu2024mimo,xu2024isac,yao2025optimal,yao2024gc,liu2024ris,hou2023optimal,hou2025base, zheng2025beyond,attiah2024uplink,liu2025mimo} have studied the transmit signal optimization for multi-antenna sensing or ISAC systems. For example, \cite{xu2023radar,xu2024mimo} first characterized the sensing PCRB for multiple-input multiple-output (MIMO) ISAC systems, and derived the optimal transmit covariance matrix for striking the optimal balance between PCRB and communication rate, where various key properties on its rank were derived. Moreover, a novel \emph{probability-dependent power focusing} effect was observed with the optimized design. Furthermore, \cite{xu2024isac,yao2025optimal} derived the optimal transmit beamforming for single-user single-target ISAC systems and multi-user multi-target ISAC systems with potential dedicated sensing beams, and derived various tight bounds on the number of sensing beams needed. \cite{hou2023optimal} devised the optimal transmit beamforming for a secure ISAC system, where dedicated sensing beams also served as artificial noise beams for enhancing secrecy. \looseness=-1

Prior studies on the transmit signal design for ISAC exploiting distribution information have focused on the case with \emph{fully-digital arrays} at the base station (BS) transmitter and the BS receiver, to unveil the fundamental new design principles brought by the prior information. However, to reduce hardware cost and energy consumption, \emph{hybrid analog-digital arrays} with \emph{limited radio frequency (RF) chains} may be employed at the BS, especially in future 6G networks with large-scale antenna arrays operating over millimeter-wave (mmWave) or higher frequency. This necessitates hybrid beamforming design under challenging non-convex unit-modulus constraints on the analog beamforming matrices \cite{el2014spatially,yu2016alternating,sohrabi2016hybrid,sohrabi2017hybrid}.
In ISAC systems with \emph{hybrid transmit array} and \emph{fully-digital receive array} at the BS, the transmit hybrid beamforming optimization has been studied in \cite{liu2019hybrid,hu2022low,nguyen2024joint,wei2025precoding,wang2022partially}, with various sensing performance metrics such as beampattern mismatch \cite{liu2019hybrid,hu2022low,nguyen2024joint}, sensing mutual information \cite{wei2025precoding}, and CRB \cite{wang2022partially}; while \cite{liu2020joint1} evaluated the performance of a heuristic receive analog beamforming design in a tracking scenario. \looseness=-1

Despite the above studies, several critical problems in hybrid beamforming for ISAC remain open. Firstly, how to design the hybrid beamforming for \emph{unknown} and \emph{random} parameters with heterogeneous probability densities, where limited RF chains constrain the beamforming ``resolution'' for both sensing and communication? Secondly, when the BS receiver is also equipped with a \emph{hybrid analog-digital array}, the hybrid signal processing at the receiver will determine the sensing performance, while that at the transmitter will affect both the sensing performance and the communication performance. How to \emph{jointly optimize} the hybrid signal processing at the transceivers is a new challenge. Finally, with a given budget on the number of RF chains at the BS, it is unclear whether it is desirable to allocate more RF chains to the BS transmitter or the BS receiver for ISAC, which is an interesting new problem.\looseness=-1

\begin{figure}[t]
	\centering
	\includegraphics[width=0.48\textwidth]{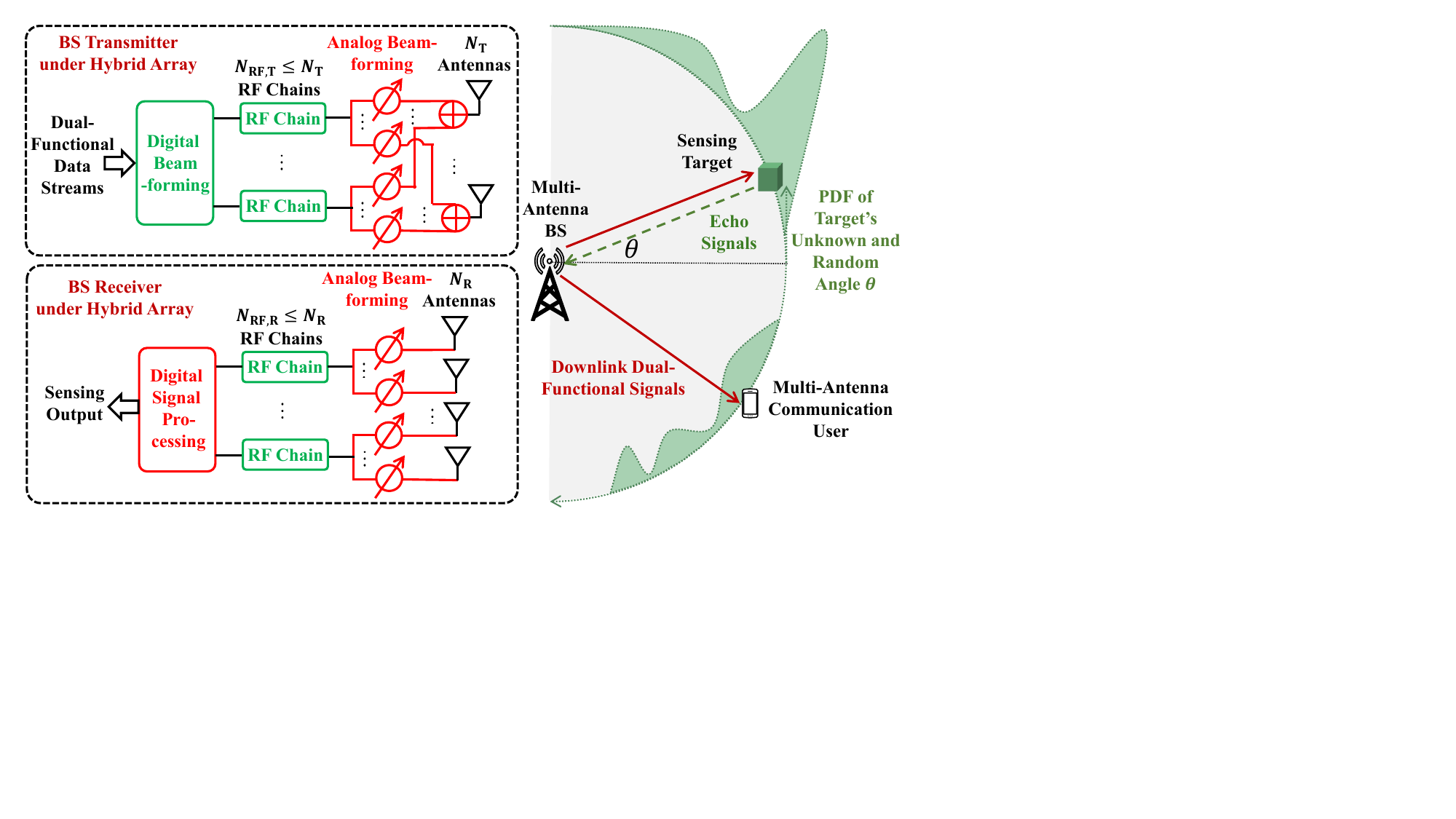}
	\vspace{-4mm}
	\caption{Illustration of MIMO ISAC under hybrid BS transceiver arrays.}
	\label{system model}
\end{figure}

Motivated by the above, this paper aims to make the first attempt to jointly design the hybrid signal processing at the BS transceivers, in the challenging scenario where the parameters to be sensed are unknown and random. We primarily consider a MIMO ISAC system with a fully-connected hybrid array at the BS transmitter and a partially-connected hybrid array at the BS receiver, as illustrated in Fig. \ref{system model}, due to its low hardware complexity and energy consumption \cite{el2014spatially},  which is particularly suitable for MIMO ISAC since the number of BS receive antennas should be no smaller than that of the BS transmit antennas to avoid information loss of sensing target \cite{liu2022cramer};\footnote{Under this architecture, the receiver noise remains white after analog combining, which can be exploited for the hybrid beamforming design and shed light on the design under the fully-connected architecture. \looseness=-1}
while the extension to the case with a fully-connected hybrid array at the receiver is presented in Section \ref{sec:extension}.
The BS aims to simultaneously communicate with a multi-antenna user equipped with a fully-digital array and sense the \emph{unknown} and \emph{random} location information of a target exploiting its prior distribution information. Our main contributions are summarized as follows.\looseness=-1

\begin{itemize}
    \item Firstly, we characterize the sensing performance under hybrid arrays at both the BS transmitter and the BS receiver by deriving the PCRB of the MSE. Different from the fully-digital case where the PCRB is \emph{independent} of the BS receiver signal processing since it is a global MSE lower bound for any unbiased estimator, the PCRB under hybrid receive array is \emph{dependent} on the \emph{receive analog beamforming}, since digital baseband signal processing for sensing can only be performed after analog beamforming. 
    \item Then, we consider a MIMO sensing-only system and jointly optimize the transmit hybrid beamforming and receive analog beamforming for PCRB minimization. Despite the non-convex unit-modulus constraints in the analog beamforming matrices and the complex PCRB expression, we derive the \emph{closed-form optimal solution} to each element in the transmit/receive analog beamforming matrices. Based on this, we develop an efficient alternating optimization (AO) based algorithm which is guaranteed to converge to a Karush-Kuhn-Tucker (KKT) point. \looseness=-1 
    \item Next, we study the joint transmit hybrid beamforming and receive analog beamforming optimization in MIMO ISAC, to minimize the sensing PCRB subject to a constraint on the MIMO communication rate. Despite the non-convexity of the problem, we devise an AO-based framework which finds a high-quality solution in polynomial time by leveraging the weighted minimum mean-squared error (WMMSE) transformation and the feasible point pursuit successive convex approximation (FPP-SCA) technique.
    \item Moreover, we extend the above results for narrowband systems to wideband orthogonal frequency-division multiplexing (OFDM) systems. In this case, a new challenge lies in the lack of frequency selectivity in the analog beamforming design, whose effect for sensing remains unknown. Motivated by this, we characterize the sensing PCRB under MIMO-OFDM, and propose an efficient AO-based algorithm for jointly optimizing the transmit digital beamforming in each sub-carrier and the common transmit/receive analog beamforming for all sub-carriers.\looseness=-1
    \item Furthermore, we demonstrate the versatility of our framework by extending it to the case under fully-connected receiver architecture via a discrete Fourier transform (DFT) codebook based method.\looseness=-1
    \item Finally, we provide extensive numerical results to evaluate the effect of limited RF chains and the performance of the proposed AO-based algorithms. It is observed that the number of receive RF chains has a higher impact on the sensing performance compared to its transmit counterpart under the considered structure. With a given budget on the total number of RF chains at the BS transceivers, it is optimal to allocate the smallest number of RF chains to the transmitter for the sensing-only system and under low-to-moderate rate requirements for MIMO ISAC and MIMO-OFDM ISAC systems; while the optimal number of RF chains allocated to the transmitter increases as the communication rate target increases due to the need of spatial multiplexing. Moreover, the proposed hybrid beamforming designs yield a probability-dependent power focusing effect, whose resolution decreases as the number of transmit/receive RF chains decreases. Lastly, the proposed AO-based algorithms outperform various benchmark schemes, and achieve close performance to the optimal fully-digital beamforming with moderate numbers of RF chains, which validates its effectiveness.\looseness=-1
\end{itemize}

\vspace{-2mm}
The remainder of this paper is organized as follows. Section \ref{sec_sys} presents the system model. Section \ref{sec_PCRB} characterizes the sensing performance under hybrid arrays. Section \ref{sec_sensing only} studies the hybrid beamforming optimization for MIMO sensing. Section \ref{sec_narrowband} studies the hybrid beamforming optimization for MIMO ISAC towards the optimal PCRB-rate trade-off, which is extended to the case of MIMO-OFDM ISAC and the case under fully-connected receiver architecture in Sections \ref{sec_OFDM} and \ref{sec:extension},  respectively. Numerical results are presented in Section \ref{sec_numerical results}. Section \ref{sec_conclusions} concludes this paper. \looseness=-1

\textit{Notations:} Vectors and matrices are denoted by boldface lower-case letters and boldface upper-case letters, respectively. $|a|$, $a^*$, $\mathrm{arg}(a)$, and $\mathfrak{Re}\{a\}$ denote the absolute value, conjugate, angle, and real part of a complex number $a$, respectively. $\mathbb{C}^{M\times N}$ and $\mathbb{R}^{M\times N}$ denote the space of $M\times N$ complex matrices and real matrices, respectively.
$\bm{I}_M$ denotes an $M\times M$ identity matrix, and $\bm{0}$ denotes an all-zero matrix with appropriate dimension. For a complex vector $\bm{a}$, $\|\bm{a}\|_1$, $\|\bm{a}\|_2$, and $a_i$ denote the $l_1$-norm, $l_2$-norm, and the $i$-th element, respectively. For a square matrix $\bm{S}$, $|\bm{S}|$, $\bm{S}^{-1}$, and $\mathrm{tr}(\bm{S})$ denote its determinant, inverse, and trace, respectively; $\bm{S}\succeq\bm{0}$ means that $\bm{S}$ is positive semi-definite. For an $M\times N$ matrix $\bm{A}$, $\bm{A}^T$, $\bm{A}^H$, $\mathrm{rank}(\bm{A})$, $\bm{A}^{\dagger}$, $\|\bm{A}\|_{\mathrm{F}}$, and $[\bm{A}]_{i,j}$ denote its transpose, conjugate transpose, rank, pseudo-inverse, Frobenius norm, and the $(i,j)$-th element, respectively. $\mathrm{diag}(\cdot)$ and $\mathrm{blkdiag}(\cdot)$ denote the diagonal and block diagonal operations, respectively. $\bm{A}\otimes\bm{B}$  denotes the Kronecker product of $\bm{A}$ and $\bm{B}$. $(z)^+\triangleq\mathrm{max}(z,0)$ with $\mathrm{max}(a,b)$ denoting the maximum between $a$ and $b$. The distribution of a circularly symmetric complex Gaussian (CSCG) random vector with mean $\bm{0}$ and variance $\bm{\Sigma}$ is denoted by $\mathcal{CN}(\bm{0},\bm{\Sigma})$, and $\sim$ means ``distributed as''. 
$\mathbb{E}[\cdot]$ denotes the statistical expectation. $\mathbb{N}[\cdot]$ denotes the operation of extracting non-zero elements. $\mathcal{O}(\cdot)$ denotes the standard big-O notation. $\dot{f}(\cdot)$ denotes the derivative of $f(\cdot)$. \looseness=-1

\vspace{-3mm}
\section{System Model}\label{sec_sys}
\vspace{-2mm}
We consider a narrowband MIMO ISAC system which consists of a multi-antenna BS equipped with $N_{\mathrm{T}}\geq 1$ transmit antennas and $N_{\mathrm{R}}\geq 1$ co-located receive antennas, a multi-antenna user equipped with $N_{\mathrm{U}}\geq 1$ antennas, and a point target whose \emph{unknown} and \emph{random} location information needs to be sensed. The BS sends dual-functional signals to communicate with the user and estimate the location information of the target via the echo signals reflected by the target and received back at the BS receive antennas. Specifically, we aim to sense the target's angular location $\theta$ with respect to the BS, as illustrated in Fig. \ref{system model}. The PDF of $\theta$ denoted by $p_{\Theta}(\theta)$ is assumed to be known \emph{a priori} based on target appearance pattern or historic data \cite{xu2023radar,xu2024mimo,xu2024isac,yao2025optimal,yao2024gc,liu2024ris,hou2023optimal,hou2025base, zheng2025beyond,attiah2024uplink,liu2025mimo}, which will be exploited in the sensing of $\theta$.\looseness=-1

We focus on a challenging case where both the BS transmitter and the BS receiver employ a \emph{hybrid analog-digital} array structure with $N_{\mathrm{RF},\mathrm{T}}< N_{\mathrm{T}}$ and $N_{\mathrm{RF},\mathrm{R}}< N_{\mathrm{R}}$ RF chains, respectively, as illustrated in Fig. \ref{system model}, which is cost-effective and particularly suitable under mmWave or higher frequency bands and/or with large-scale BS antenna arrays. Specifically, the BS transmitter employs a fully-connected hybrid analog-digital array, and the BS receiver employs a partially-connected hybrid analog-digital array where $\frac{N_{\mathrm{R}}}{N_{\mathrm{RF},\mathrm{R}}}$ is an integer. By further considering a linear beamforming model, we let $\bm{V}_{\mathrm{RF}}\in \mathbb{C}^{N_{\mathrm{T}}\times N_{\mathrm{RF},\mathrm{T}}}$ with $|[\bm{V}_{\mathrm{RF}}]_{i,j}|=1, \forall i,j$ denote the fully-connected analog beamforming matrix and $\bm{V}_{\mathrm{BB}}\in \mathbb{C}^{N_{\mathrm{RF},\mathrm{T}}\times N_{\mathrm{S}}}$ denote the digital beamforming matrix at the BS transmitter,  where $N_{\mathrm{S}}$ denotes the number of data streams with $N_{\mathrm{S}}\leq N_{\mathrm{RF},\mathrm{T}}$. Furthermore, let $\bm{W}_{\mathrm{RF}}^H\in \mathbb{C}^{N_{\mathrm{RF},\mathrm{R}}\times N_{\mathrm{R}}}$ denote the  partially-connected analog beamforming matrix at the BS receiver, which satisfies $\bm{W}_{\mathrm{RF}}^H\bm{W}_{\mathrm{RF}}=\frac{N_{\mathrm{R}}}{N_{\mathrm{RF},\mathrm{R}}}\bm{I}_{N_{\mathrm{RF},\mathrm{R}}}$ \cite{sohrabi2017hybrid} via
\begin{equation}\label{WRF_constraint}
    |[\bm{W}_{\mathrm{RF}}^H]_{i,j}|=\begin{cases} 1, \text{if}\ (i-1)\frac{N_{\mathrm{R}}}{N_{\mathrm{RF},\mathrm{R}}}+1\leq j \leq i\frac{N_{\mathrm{R}}}{N_{\mathrm{RF},\mathrm{R}}}\\0, \text{otherwise}.
\end{cases}
\end{equation}
As the number of antennas at the user is generally small or moderate, we consider a fully-digital array at the user receiver. Note that all the results in this paper are also directly applicable to the case with fully-digital transmitter and/or receiver at the BS, by setting $\bm{V}_{\mathrm{RF}}=\bm{I}_{N_{\mathrm{T}}}$ and/or $\bm{W}_{\mathrm{RF}}^H=\bm{I}_{N_\mathrm{R}}$.

\vspace{-5mm}
\subsection{MIMO Communication Model under Hybrid Arrays}
\vspace{-1mm}
Consider a quasi-static block-fading frequency-flat channel model between the BS and the user, where the channel remains constant within each coherence block consisting of $L_\mathrm{C}$ symbol intervals. Let $\bm{H}\in \mathbb{C}^{N_{\mathrm{U}}\times N_{\mathrm{T}}}$ denote the channel matrix from the BS to the user, which is assumed to be perfectly known at the BS and the user via channel estimation techniques (see, e.g., \cite{alkhateeb2014channel}). We focus on $L\leq L_{\mathrm{C}}$ symbol intervals during which both the communication channel and sensing target remain static.\looseness=-1

Let $\bm{s}_l\in \mathbb{C}^{N_{\mathrm{S}}\times 1}$ denote the data symbol vector in the $l$-th interval, with $\bm{s}_l\sim \mathcal{CN}(\bm{0},\bm{I}_{N_\mathrm{S}})$. The baseband equivalent transmit signal vector in each $l$-th interval is given by \looseness=-1
    \begin{equation}
		\bm{x}_l=\bm{V}_{\mathrm{RF}}\bm{V}_{\mathrm{BB}}\bm{s}_l,\quad l=1,...,L.
    \end{equation}
The transmit covariance matrix is thus given by
\begin{equation}
	\tilde{\bm{R}}_X=\mathbb{E}[\bm{x}_l\bm{x}_l^H]=\bm{V}_{\mathrm{RF}}\bm{V}_{\mathrm{BB}}\bm{V}_{\mathrm{BB}}^H\bm{V}_{\mathrm{RF}}^H=\bm{V}_{\mathrm{RF}}\bm{R}_{\mathrm{BB}}\bm{V}_{\mathrm{RF}}^H,
\end{equation}
where $\bm{R}_{\mathrm{BB}}=\bm{V}_{\mathrm{BB}}\bm{V}_{\mathrm{BB}}^H$ denotes the covariance matrix for transmit digital beamforming. Let $P$ denote the transmit power budget, which yields $\mathrm{tr}(\tilde{\bm{R}}_X)=\mathrm{tr}(\bm{V}_{\mathrm{RF}}\bm{V}_{\mathrm{BB}}\bm{V}_{\mathrm{BB}}^H\bm{V}_{\mathrm{RF}}^H)\leq P$. 
The received signal vector at the user in each $l$-th interval is given by \looseness=-1
    \begin{equation}
		\bm{y}_l^{\mathrm{C}}=\bm{H}\bm{V}_{\mathrm{RF}}\bm{V}_{\mathrm{BB}}\bm{s}_l+\bm{n}_l^{\mathrm{C}},  \quad l=1,...,L,
    \end{equation}
where $\bm{n}_l^{\mathrm{C}}\sim\mathcal{CN}(\bm{0}, \sigma_{\mathrm{C}}^2\bm{I}_{N_{\mathrm{U}}})$ denotes the noise vector at the user receiver, with $\sigma_{\mathrm{C}}^2$ denoting the average noise power. The achievable rate for the communication user is then given by
    \begin{equation}\label{rate_narrow}
		R=\log\bigg|\bm{I}_{N_{\mathrm{U}}}+\frac{\bm{H}\bm{V}_{\mathrm{RF}}\bm{R}_{\mathrm{BB}}\bm{V}_{\mathrm{RF}}^H\bm{H}^H}{\textcolor{black}{\sigma_{\mathrm{C}}^2}}\bigg|
    \end{equation}
in nats per second per Hertz (nps/Hz).

\vspace{-5mm}
\subsection{MIMO Sensing Model under Hybrid Arrays}
\vspace{-1mm}
By sending downlink signals, the BS can also sense $\theta$ based on the received echo signals reflected by the target and the prior distribution information about $\theta$. We consider a line-of-sight (LoS) channel between the BS and the target. The equivalent MIMO channel from the BS transmitter to the BS receiver via target reflection is given by $\alpha\bm{b}(\theta)\bm{a}^H(\theta)$.
Specifically, $\alpha\triangleq \frac{\beta_0}{r^2}\psi=\alpha_{\mathrm{R}}+j\alpha_{\mathrm{I}}\in \mathbb{C}$ denotes the reflection coefficient, which contains both the round-trip channel gain $\frac{\beta_0}{r^2}$ with $\beta_0$ denoting the reference channel power at distance $1$ m and $r$ denoting the BS-target distance in m, as well as the radar cross-section (RCS) coefficient $\psi\in\mathbb{C}$. Note that $\alpha$ is generally unknown, while its distribution can be known \emph{a priori} based on target properties. Moreover, $\bm{a}(\theta)\in \mathbb{C}^{N_{\mathrm{T}}\times 1}$ and $\bm{b}(\theta)\in \mathbb{C}^{N_{\mathrm{R}}\times 1}$ denote the steering vectors of the transmit and receive antennas, respectively, in which $\theta$ is the only unknown parameter.
The received echo signal vector in each $l$-th symbol interval is given by \looseness=-1
    \begin{align}
            \bm{y}_l^{\mathrm{S}}&=\alpha\bm{b}(\theta)\bm{a}^H(\theta)\bm{V}_{\mathrm{RF}}\bm{V}_{\mathrm{BB}}\bm{s}_l+\bm{n}_l^{\mathrm{S}},   \quad l=1,...,L,
    \end{align}
where $\bm{n}_l^{\mathrm{S}}\sim\mathcal{CN}(\bm{0}, \sigma_{\mathrm{S}}^2\bm{I}_{N_{\mathrm{R}}})$ denotes the CSCG noise vector at the BS receiver, with $\sigma_{\mathrm{S}}^2$ denoting the average noise power. 

Note that different from sensing under a fully-digital array structure which is based on $\bm{y}_l^{\mathrm{S}}$'s, only the analog-beamformed version of $\bm{y}_l^{\mathrm{S}}$'s can be used for baseband processing and consequently sensing under the considered hybrid array structure. The analog beamforming output (or the ``effective received signal'') at the BS receiver in each $l$-th symbol interval is given by \looseness=-1
    \begin{align}
            \tilde{\bm{y}}_l^{\mathrm{S}}&=\bm{W}_{\mathrm{RF}}^H\alpha\bm{b}(\theta)\bm{a}^H(\theta)\bm{V}_{\mathrm{RF}}\bm{V}_{\mathrm{BB}}\bm{s}_l+\bm{W}_{\mathrm{RF}}^H\bm{n}_l^{\mathrm{S}} \nonumber\\
            &=\bm{W}_{\mathrm{RF}}^H\alpha\bm{b}(\theta)\bm{a}^H(\theta)\bm{V}_{\mathrm{RF}}\bm{V}_{\mathrm{BB}}\bm{s}_l+\bm{z}_l,  \quad l=1,...,L.
    \end{align}
The effective noise $\bm{z}_l$ follows $\bm{z}_l\sim\mathcal{CN}(\bm{0}, \sigma_{\mathrm{S}}^2\bm{W}_{\mathrm{RF}}^H\bm{W}_{\mathrm{RF}})$.
The collection of received signal vectors available for sensing over $L$ symbol intervals can be expressed as
    \begin{align}\label{Y1}
            \tilde{\bm{Y}}^{\mathrm{S}}&=[\tilde{\bm{y}}_1^{\mathrm{S}},...,\tilde{\bm{y}}_L^{\mathrm{S}}]=\bm{W}_{\mathrm{RF}}^H\alpha\bm{b}(\theta)\bm{a}^H(\theta)\bm{V}_{\mathrm{RF}}\bm{V}_{\mathrm{BB}}\bm{S}+\bm{W}_{\mathrm{RF}}^H\bm{N} \nonumber\\
            &=\bm{W}_{\mathrm{RF}}^H\alpha\bm{b}(\theta)\bm{a}^H(\theta)\bm{V}_{\mathrm{RF}}\bm{V}_{\mathrm{BB}}\bm{S}+\bm{Z},
    \end{align}
where $\bm{N}=[\bm{n}_1^{\mathrm{S}},...,\bm{n}_L^{\mathrm{S}}]$ and $\bm{Z}=[\bm{z}_1,...,\bm{z}_L]$. Since $\alpha$ is also unknown in $\tilde{\bm{Y}}^{\mathrm{S}}$, $\theta$ and $\alpha$ need to be jointly sensed based on $\tilde{\bm{Y}}^{\mathrm{S}}$ and prior distribution information of $\theta$ and $\alpha$.

\vspace{-1mm}
Notice from (\ref{Y1}) and (\ref{rate_narrow}) that both the received echo signals for sensing and the achievable rate for communication depend on the analog and digital beamforming matrices at the BS transmitter; while the received echo signals are also dependent on the analog beamforming matrix at the BS receiver. Particularly, the number of transmit RF chains at the BS affects the rank of the transmit covariance matrix, thereby affecting the total number of dual-functional data streams for communication and sensing; on the other hand, the number of receive RF chains at the BS affects the rank of the covariance matrix of the ``effective received signals'' in $\tilde{\bm{y}}_l^{\mathrm{S}}$'s, thereby affecting the ``resolution'' of the received echo signals and consequently the sensing accuracy. Therefore, the BS hybrid beamforming at the transceivers needs to be judiciously optimized such that the limited RF chains can be optimally shared by both sensing and communication functions based on the prior distribution information about the target. In the next section, we will characterize the sensing performance under the hybrid array structure exploiting prior distribution information.

\vspace{-4mm}
\section{Characterization of Sensing Performance under Hybrid Arrays Exploiting Prior Information}\label{sec_PCRB}
\vspace{-2mm}

In this section, we aim to characterize the sensing performance for $\theta$ exploiting prior information under hybrid analog-digital arrays at the BS transceivers. Since the sensing MSE is difficult to express analytically, we adopt the PCRB as the performance metric, which is a lower bound of the MSE.

Define $\bm{\zeta}=[\theta, \alpha_{\mathrm{R}}, \alpha_{\mathrm{I}}]^T$ as the collection of all the unknown (real) parameters, which need to be jointly sensed to obtain an accurate estimate of $\theta$. We assume that all the parameters in $\bm{\zeta}$ are independent of each other. Let $p_{\alpha_{\mathrm{R}},\alpha_{\mathrm{I}}}(\alpha_{\mathrm{R}},\alpha_{\mathrm{I}})$ denote the PDF for $\alpha$ and assume $\alpha$ is independent of $\theta$. The PDF of $\bm{\zeta}$ is given by $p_Z(\bm{\zeta})=p_{\Theta}(\theta)p_{\alpha_{\mathrm{R}},\alpha_{\mathrm{I}}}(\alpha_{\mathrm{R}},\alpha_{\mathrm{I}})$.\footnote{We consider a mild condition of $\iint\alpha_{\mathrm{R}}p_{\alpha_{\mathrm{R}},\alpha_{\mathrm{I}}}(\alpha_{\mathrm{R}},\alpha_{\mathrm{I}})d\alpha_{\mathrm{R}}d\alpha_{\mathrm{I}}=\iint\alpha_{\mathrm{I}}p_{\alpha_{\mathrm{R}},\alpha_{\mathrm{I}}}(\alpha_{\mathrm{R}},\alpha_{\mathrm{I}})d\alpha_{\mathrm{R}}d\alpha_{\mathrm{I}}=0$, which holds for various random variables (RVs), including but not limited to all proper RVs such as CSCG RVs.\looseness=-1} The posterior Fisher information matrix (PFIM) for sensing $\bm{\zeta}$ is given by $ \bm{F}=\bm{F}_{\mathrm{O}}+\bm{F}_{\mathrm{P}}$ \cite{shen2010fundamental}, 
where $\bm{F}_{\mathrm{O}}\in\mathbb{R}^{3\times 3}$ denotes the PFIM from the observation in (\ref{Y1}), and $\bm{F}_{\mathrm{P}}\in\mathbb{R}^{3\times 3}$ denotes the PFIM from prior information in $p_Z(\bm{\zeta})$. For ease of exposition, we vectorize (\ref{Y1}), which yields
$\bm{y}=\mathrm{vec}(\tilde{\bm{Y}}^{\mathrm{S}})=\bm{u}+\bm{z}\sim\mathcal{CN}(\bm{u},\bm{C}_y)$ with $\bm{u}=\alpha\mathrm{vec}(\bm{W}_{\mathrm{RF}}^H\bm{b}(\theta)\bm{a}^H(\theta)\bm{V}_{\mathrm{RF}}\bm{V}_{\mathrm{BB}}\bm{S})$, $\bm{z}=\mathrm{vec}(\bm{Z})$, and $\bm{C}_y=\sigma_{\mathrm{S}}^2(\bm{I}_L\otimes\bm{W}_{\mathrm{RF}}^H)(\bm{I}_L\otimes\bm{W}_{\mathrm{RF}}^H)^H=\frac{N_{\mathrm{R}}}{N_{\mathrm{RF},\mathrm{R}}}\sigma_{\mathrm{S}}^2\bm{I}_{N_{\mathrm{RF},\mathrm{R}}L}$.
The entry in the $i$-th row and $j$-th column of $\bm{F}_{\mathrm{O}}$ is given by \cite{kay1993fundamentals}
    \begin{equation}\label{F_O}
        [\bm{F}_{\mathrm{O}}]_{i,j}=2\mathbb{E}_{\bm{y},\bm{\zeta}}\bigg[\mathfrak{Re}\bigg\{\bigg(\frac{\partial\bm{u}}{\partial\zeta_i}\bigg)^H\bm{C}_y^{-1}\frac{\partial\bm{u}}{\partial\zeta_j}\bigg\}\bigg],\quad  \forall i,j.
    \end{equation}
Based on (\ref{F_O}), the PFIM $\bm{F}_{\mathrm{O}}$ from observation can be partitioned as $\bm{F}_{\mathrm{O}}=\begin{bmatrix}
            J_{\theta\theta} & \bm{J}_{\theta\alpha} \\
            \bm{J}_{\theta\alpha}^H & \bm{J}_{\alpha\alpha}
        \end{bmatrix}$.
Specifically, we assume that $L$ is sufficiently large such that the sample covariance matrix $\bm{R}_X=\frac{1}{L}\sum_{l=1}^L\bm{x}_l\bm{x}_l^H$ can
be accurately approximated by the transmit covariance matrix $\tilde{\bm{R}}_X=\bm{V}_{\mathrm{RF}}\bm{R}_{\mathrm{BB}}\bm{V}_{\mathrm{RF}}^H$ \cite{hua2024mimo}. 
Each block in $\bm{F}_{\mathrm{O}}$ can be then derived as
\vspace{1mm}
    \begin{align}\label{J_thetatheta}
            \!\!\!\!\!\!\!\!\!\!\! J_{\theta\theta}
            =&\frac{2N_{\mathrm{RF},\mathrm{R}}L}{N_{\mathrm{R}}\sigma_{\mathrm{S}}^2}\mathbb{E}_{\zeta}\Big[\mathfrak{Re}\Big\{
            (\alpha_{\mathrm{R}}^2+\alpha_{\mathrm{I}}^2)\nonumber\\
            &\times\mathrm{tr}(\Dot{\bm{M}}^H(\theta)\bm{W}_{\mathrm{RF}}\bm{W}_{\mathrm{RF}}^H\Dot{\bm{M}}(\theta)\bm{V}_{\mathrm{RF}}\bm{R}_{\mathrm{BB}}\bm{V}_{\mathrm{RF}}^H)\Big\}
            \Big] \notag\\
            =&\frac{2N_{\mathrm{RF},\mathrm{R}}L\gamma}{N_{\mathrm{R}}\sigma_{\mathrm{S}}^2}\mathrm{tr}(\bm{A}_1(\bm{W}_{\mathrm{RF}}^H)\bm{V}_{\mathrm{RF}}\bm{R}_{\mathrm{BB}}\bm{V}_{\mathrm{RF}}^H),
    \end{align}
    \begin{align}\label{J_thetaalpha}
            \bm{J}_{\theta\alpha}
            =&\frac{2N_{\mathrm{RF},\mathrm{R}}L}{N_{\mathrm{R}}\sigma_{\mathrm{S}}^2}\mathbb{E}_{\zeta}\Big[
            \mathfrak{Re}\big\{\alpha^*\nonumber\\
            &\times\mathrm{tr}(\Dot{\bm{M}}^H(\theta)\bm{W}_{\mathrm{RF}}\bm{W}_{\mathrm{RF}}^H\bm{M}(\theta)\bm{V}_{\mathrm{RF}}\bm{R}_{\mathrm{BB}}\bm{V}_{\mathrm{RF}}^H)[1,j]\big\}
            \Big] \notag\\
            =&\bm{0},
    \end{align}
    \begin{align}\label{J_alphaalpha}
            \bm{J}_{\alpha\alpha}
            =&\frac{2N_{\mathrm{RF},\mathrm{R}}L}{N_{\mathrm{R}}\sigma_{\mathrm{S}}^2}\mathfrak{Re}\Big\{
            \begin{bmatrix}
                1 & j\\
                -j& 1
            \end{bmatrix}\nonumber\\
            &\times\mathrm{tr}\Big(\mathbb{E}_{\theta}\Big[\bm{M}^H(\theta)\bm{W}_{\mathrm{RF}}\bm{W}_{\mathrm{RF}}^H\bm{M}(\theta)\Big]\bm{V}_{\mathrm{RF}}\bm{R}_{\mathrm{BB}}\bm{V}_{\mathrm{RF}}^H\Big)\Big\}\nonumber\\
            =&\frac{2N_{\mathrm{RF},\mathrm{R}}L}{N_{\mathrm{R}}\sigma_{\mathrm{S}}^2}\mathrm{tr}(\bm{A}_2(\bm{W}_{\mathrm{RF}}^H)\bm{V}_{\mathrm{RF}}\bm{R}_{\mathrm{BB}}\bm{V}_{\mathrm{RF}}^H)\bm{I}_2,
    \end{align}
where $\gamma\triangleq\iint(\alpha_{\mathrm{R}}^2+\alpha_{\mathrm{I}}^2)p_{\alpha_{\mathrm{R}},\alpha_{\mathrm{I}}}(\alpha_{\mathrm{R}},\alpha_{\mathrm{I}})d\alpha_{\mathrm{R}}d\alpha_{\mathrm{I}}$,
    \begin{align}
        \bm{A}_1(\bm{W}_{\mathrm{RF}}^H)&\!=\!\int\Dot{\bm{M}}^H(\theta)\bm{W}_{\mathrm{RF}}\bm{W}_{\mathrm{RF}}^H\Dot{\bm{M}}(\theta)p_{\Theta}(\theta)d\theta\succeq\bm{0}, \label{eq_A1}\\
        \bm{A}_2(\bm{W}_{\mathrm{RF}}^H)&\!=\!\int\bm{M}^H(\theta)\bm{W}_{\mathrm{RF}}\bm{W}_{\mathrm{RF}}^H\bm{M}(\theta)p_{\Theta}(\theta)d\theta\succeq\bm{0},
    \end{align}
and $\bm{M}(\theta)\triangleq\bm{b}(\theta)\bm{a}^H(\theta)$.
Moreover, $\bm{F}_{\mathrm{P}}$ can be derived as
    \begin{align}
        \bm{F}_{\mathrm{P}}= &\mathbb{E}_{\zeta}\Big[
        \frac{\partial\ln(p_Z(\bm{\zeta}))}{\partial\bm{\zeta}}\Big(\frac{\partial\ln(p_Z(\bm{\zeta}))}{\partial\bm{\zeta}}\Big)^H
        \Big] \nonumber\\
        =&\bigg[\mathbb{E}_\theta\Big[\Big(\frac{\partial\ln(p_{\Theta}(\theta))}{\partial\theta}\Big)^2\Big],\bm{0};\bm{0},\bm{F}^{\alpha\alpha}_{\mathrm{P}}\bigg],
    \end{align} 
where $[\bm{F}^{\alpha\alpha}_{\mathrm{P}}]_{m,n}\!\!=\!\mathbb{E}_{\alpha_{\mathrm{R}},\alpha_{\mathrm{I}}}\bigg[\frac{\partial \ln(p_{\alpha_{\mathrm{R}},\alpha_{\mathrm{I}}}(\alpha_{\mathrm{R}},\alpha_{\mathrm{I}}))}{\partial \bar{\alpha}_m }\frac{\partial \ln(p_{\alpha_{\mathrm{R}},\alpha_{\mathrm{I}}}(\alpha_{\mathrm{R}},\alpha_{\mathrm{I}}))}{\partial \bar{\alpha}_n }\!\bigg]$.
The overall PCRB for the MSE matrix of estimating $\bm{\zeta}$ is given by $\bm{F}^{-1}$. The PCRB for the MSE in estimating the desired sensing parameter $\theta$ is given by
\vspace{-1mm}
    \begin{align}\label{PCRB_theta}
        \mathrm{PCRB}_{\theta}&=[\bm{F}^{-1}]_{1,1}=1\bigg/\bigg(\mathbb{E}_\theta\Big[\Big(\frac{\partial\ln(p_{\Theta}(\theta))}{\partial\theta}\Big)^2\Big]\nonumber\\
        &+\frac{2N_{\mathrm{RF},\mathrm{R}}L\gamma}{N_{\mathrm{R}}\sigma_{\mathrm{S}}^2}\mathrm{tr}(\bm{A}_1(\bm{W}_{\mathrm{RF}}^H)\bm{V}_{\mathrm{RF}}\bm{R}_{\mathrm{BB}}\bm{V}_{\mathrm{RF}}^H)
        \bigg).
    \end{align}
It is worth noting that the $\mathrm{PCRB}_{\theta}$ is a global lower bound for the MSE under any digital baseband processing at the BS receiver, thereby being independent of the latter.

Note that both the communication rate $R$ in (\ref{rate_narrow}) and $\mathrm{PCRB}_{\theta}$ in (\ref{PCRB_theta}) are functions of the transmit analog and digital beamforming matrices in $\bm{V}_{\mathrm{RF}}$ and $\bm{R}_{\mathrm{BB}}$. Moreover, $\mathrm{PCRB}_{\theta}$ is also a function of the receive analog beamforming matrix $\bm{W}_{\mathrm{RF}}^H$ as well as the prior distribution information.

\vspace{-5mm}
\section{Hybrid Beamforming for MIMO Sensing}\label{sec_sensing only}
\vspace{-1mm}
In this section, we aim to draw fundamental insights into the effect of limited RF chains on the sensing performance. To this end, we focus on a system with only the sensing function.

\vspace{-4mm}
\subsection{Problem Formulation}
\vspace{-1mm}
We aim to jointly optimize the transmit hybrid beamforming and receive analog beamforming to minimize the sensing PCRB. The optimization problem is formulated as
    \begin{align}
			\mbox{(P0)}\quad \underset{\bm{V}_\mathrm{RF},\bm{R}_{\mathrm{BB}}\succeq\bm{0},\bm{W}_{\mathrm{RF}}^H:(\ref{WRF_constraint})}{\min}& \mathrm{PCRB}_{\theta} \label{P0_obj}\\
			\mbox{s.t.}\qquad\quad\ & \mathrm{tr}(\bm{V}_{\mathrm{RF}}\bm{R}_{\mathrm{BB}}\bm{V}_{\mathrm{RF}}^H)\leq P \label{P0_power constraint}\\
			& |[\bm{V}_{\mathrm{RF}}]_{m,n}|=1,\quad \forall m,n. \label{P0_unit-modulus constraint1}
    \end{align}
Note that (P0) is a non-convex optimization problem due to the unit-modulus constraints in (\ref{WRF_constraint}) and (\ref{P0_unit-modulus constraint1}). To tackle this problem, we first present its optimal solution under special array structures, and then present a general algorithm for (P0).

\vspace{-4mm}
\subsection{Optimal Solution under Special Array Structures}\label{subsec_sening_B}
\vspace{-1mm}
\subsubsection{Case I} \emph{Fully-digital transceiver arrays.}\label{subsubsectionCaseI} In this case, we have $N_{\mathrm{RF},\mathrm{T}}=N_{\mathrm{T}}$, $N_{\mathrm{RF},\mathrm{R}}=N_{\mathrm{R}}$, as well as $\bm{V}_{\mathrm{RF}}=\bm{I}_{N_{\mathrm{T}}}$ and $\bm{W}_{\mathrm{RF}}^H=\bm{I}_{N_\mathrm{R}}$. (P0) is thus equivalent to 
		\begin{align}
			\mbox{(P0-FD)}\quad \underset{\bm{R}_{\mathrm{BB}}\succeq\bm{0}}{\max} \quad& \mathrm{tr}(\bm{A}\bm{R}_{\mathrm{BB}}) \label{P0I_obj}\\
			\mbox{s.t.}\quad\quad & \mathrm{tr}(\bm{R}_{\mathrm{BB}})\leq P, \label{P0I_power constraint}
		\end{align}
where $\bm{A}=\bm{A}_1(\bm{I}_{N_\mathrm{R}})=\int\Dot{\bm{M}}^H(\theta)\Dot{\bm{M}}(\theta)p_{\Theta}(\theta)d\theta$. (P0-FD) is a convex semi-definite program (SDP) for which an optimal solution can be shown to be given by $\bm{R}_{\mathrm{BB}}^{\star}=P\bm{q}_1\bm{q}_1^H$, where $\bm{q}_1\in\mathbb{C}^{N_{\mathrm{T}}\times 1}$ denotes the eigenvector corresponding to the strongest eigenvalue of $\bm{A}$. Note that this indicates at most $N_{\mathrm{S}}=1$ beam is needed for optimal sensing in the fully-digital case. \looseness=-1

\subsubsection{Case II} \label{subsubsectionCaseII}
\emph{Hybrid transmit array, fully-digital receive array.} This case corresponds to $\bm{W}_{\mathrm{RF}}^H=\bm{I}_{N_{\mathrm{R}}}$. The problem is formulated as
		\begin{align}
			\mbox{(P0-FDR)}\underset{\bm{V}_\mathrm{RF}:(\ref{P0_power constraint}),(\ref{P0_unit-modulus constraint1}),\bm{R}_{\mathrm{BB}}\succeq\bm{0}}{\max}& \mathrm{tr}(\bm{A}\bm{V}_{\mathrm{RF}}\bm{R}_{\mathrm{BB}}\bm{V}_{\mathrm{RF}}^H). \label{P0II_obj}
		\end{align}
(P0-FDR) is a non-convex problem. However, inspired by the results in Case I, we have the following proposition. \looseness=-1
\begin{proposition}\label{prop_solution1}
    When $N_{\mathrm{RF},\mathrm{T}}\geq 2$, hybrid transmit beamforming with fully-digital receive array can achieve the same PCRB performance as the optimal fully-digital beamforming.
\end{proposition}
\begin{IEEEproof}
    When the number of transmit RF chains is no smaller than twice the number of data streams, it was shown in \cite{zhang2005phase}, \cite{sohrabi2016hybrid} that for any $\bm{R}_{\mathrm{BB}}^\star\succeq\bm{0}$ under (\ref{P0_power constraint}) for (P0-FD), a set of $\bm{V}_{\mathrm{RF}}$ and $\bm{R}_{\mathrm{BB}}\succeq\bm{0}$ under (\ref{P0_power constraint}) and (\ref{P0_unit-modulus constraint1}) can always be found such that $\bm{V}_{\mathrm{RF}}\bm{R}_{\mathrm{BB}}\bm{V}_{\mathrm{RF}}^H=\bm{R}_{\mathrm{BB}}^{\star}$. Since sending one data stream is optimal for the fully-digital case in (P0-FD), hybrid beamforming can achieve the optimal PCRB performance with fully-digital beamforming when $N_{\mathrm{RF},\mathrm{T}}\geq 2$, $N_{\mathrm{RF},\mathrm{R}}=N_{\mathrm{R}}$.
\end{IEEEproof}
Thus, if $N_{\mathrm{RF},\mathrm{T}}\geq2$, the optimal solution to (P0-FDR) can be obtained by first solving (P0-FD) and then constructing the optimal equivalent hybrid beamforming design according to \cite{sohrabi2016hybrid}.

\vspace{-4mm}
\subsection{Proposed General AO-based Solution to (P0)}
\vspace{-1mm}

Based on the results in Section \ref{subsec_sening_B} and via exploiting the structure of the problem, in this subsection, we propose a general AO-based algorithm for (P0). Specifically, we first derive the \emph{optimal solution} to each element in the analog receive beamforming matrix. Then, for the case of $N_{\mathrm{RF},\mathrm{T}}\geq2$, the optimal solution to $\bm{V}_{\mathrm{RF}}$ and $\bm{R}_{\mathrm{BB}}$ with given $\bm{W}_{\mathrm{RF}}^H$ can be constructed in a similar manner as that for Case II in Section \ref{subsubsectionCaseII}; on the other hand, for the case of $N_{\mathrm{RF},\mathrm{T}}=1$, $\bm{R}_{\mathrm{BB}}$ is reduced to a scalar with the optimal solution given by ${R}_{\mathrm{BB}}^{\star}=\frac{P}{N_{\mathrm{T}}}$, and we derive the \emph{optimal solution} to each element in the analog transmit beamforming matrix. Based on the derived optimal solutions, we propose an AO-based algorithm to iteratively optimize the variables in (P0). The two key sub-problems are detailed and solved below.

\subsubsection{Optimal solution to each element in $\bm{W}_{\mathrm{RF}}^H$}\label{sec_Optimize_WRF}
With given $\bm{V}_{\mathrm{RF}}$ and $\bm{R}_{\mathrm{BB}}$, minimizing $\mathrm{PCRB}_{\theta}$ is equivalent to maximizing $\mathrm{tr}(\tilde{\bm{B}}\bm{W}_{\mathrm{RF}}\bm{W}_{\mathrm{RF}}^H)$, where $    \tilde{\bm{B}}=\int\Dot{\bm{M}}(\theta)\bm{V}_{\mathrm{RF}}\bm{R}_{\mathrm{BB}}\bm{V}_{\mathrm{RF}}^H\Dot{\bm{M}}^H(\theta)p_{\Theta}(\theta)d\theta$.
Since $\mathrm{tr}(\bm{A}\bm{B}\bm{C}\bm{D})=(\mathrm{vec}(\bm{D}^T))^T(\bm{C}^T\otimes\bm{A})\mathrm{vec}(\bm{B})$, $\mathrm{tr}(\tilde{\bm{B}}\bm{W}_{\mathrm{RF}}\bm{W}_{\mathrm{RF}}^H)$ can be equivalently expressed as $\mathrm{vec}(\bm{W}_{\mathrm{RF}}^H)^H(\tilde{\bm{B}}^T\!\!\otimes\!\bm{I}_{N_{\mathrm{RF},\mathrm{R}}})\mathrm{vec}(\bm{W}_{\mathrm{RF}}^H)$.
Note that the receive analog beamforming matrix has a block-diagonal structure of $\bm{W}_{\mathrm{RF}}^H=\mathrm{blkdiag}(\bm{w}_1^H,...,\bm{w}_{N_{\mathrm{RF},\mathrm{R}}}^H)$, where $\bm{w}_i\in\mathbb{C}^{M\times 1}$ with $M=\frac{N_{\mathrm{R}}}{N_{\mathrm{RF},\mathrm{R}}}$.
The non-zero elements of $\bm{W}_{\mathrm{RF}}^H$ can be extracted  as \looseness=-1
    \begin{equation}
        \bm{d}=\mathbb{N}[\mathrm{vec}(\bm{W}_{\mathrm{RF}}^H)]=[\bm{w}_1^H,...,\bm{w}_{N_{\mathrm{RF},\mathrm{R}}}^H]^T\in\mathbb{C}^{N_{\mathrm{R}}\times 1},
    \end{equation}
    where $|d_n|=1,\ n=1,...,N_{\mathrm{R}}$.
    Subsequently, we extract the elements in $\tilde{\bm{B}}^T\otimes\bm{I}_{N_{\mathrm{RF},\mathrm{R}}}$ corresponding to $\bm{d}$. Specifically, we construct $\bm{\Pi}\in\mathbb{C}^{N_{\mathrm{R}}\times N_{\mathrm{R}}}$ via
    \begin{align}
            &\bm{\Pi}((i-1)M+1:iM,(i-1)M+1:iM) \nonumber\\
            &=\tilde{\bm{B}}^T((i-1)M+1:iM,(i-1)M+1:iM),
    \end{align}
    where $i=1,...,N_{\mathrm{RF},\mathrm{R}}$. Through the above manipulations, $\mathrm{tr}(\tilde{\bm{B}}\bm{W}_{\mathrm{RF}}\bm{W}_{\mathrm{RF}}^H)$ can be equivalently expressed as $\bm{d}^H\bm{\Pi}\bm{d}=2\mathfrak{Re}\{d_n\sum_{i\neq n}^{N_{\mathrm{R}}}d_i^*\Pi_{i,n}\}+(\sum_{i\neq n}^{N_{\mathrm{R}}}\sum_{j\neq i,j\neq n}^{N_{\mathrm{R}}}d_i^*\Pi_{i,j}d_j+\mathrm{tr}(\bm{\Pi})),\forall n$. The optimal $d_n$ that maximizes $\bm{d}^H\bm{\Pi}\bm{d}$ with the other optimization variables being fixed is given by \looseness=-1
    \begin{equation}\label{wRF_solution}
        d_n^{\star}=\begin{cases}
            1, &\!\! \text{if}\, \sum_{i\neq n}^{N_\mathrm{R}}d_i\Pi_{i,n}^*=0\\
            \frac{\sum_{i\neq n}^{N_\mathrm{R}}d_i\Pi_{i,n}^*}{|\sum_{i\neq n}^{N_\mathrm{R}}d_i\Pi_{i,n}^*|}, & \!\!\text{otherwise}.
        \end{cases}
    \end{equation}
    $\bm{W}_{\mathrm{RF}}^H$ can be recovered via
    \begin{align}
			&\bm{W}_{\mathrm{RF}}^H\bigg(m,(m-1)\frac{N_{\mathrm{R}}}{N_{\mathrm{RF},\mathrm{R}}}+1:m\frac{N_{\mathrm{R}}}{N_{\mathrm{RF},\mathrm{R}}}\bigg)\\
			&=\bm{d}\bigg((m-1)\frac{N_{\mathrm{R}}}{N_{\mathrm{RF},\mathrm{R}}}\!+\!1\!:\!m\frac{N_{\mathrm{R}}}{N_{\mathrm{RF},\mathrm{R}}}\bigg), \,m=1,...,N_{\mathrm{RF},\mathrm{R}}. \nonumber
	\end{align}

\subsubsection{Optimal solution to each element in $\bm{V}_{\mathrm{RF}}$ with $N_{\mathrm{RF},\mathrm{T}}=1$}
In this case, $\bm{V}_{\mathrm{RF}}$ is reduced to a vector $\bm{v}_{\mathrm{RF}}\in \mathbb{C}^{N_{\mathrm{T}}\times 1}$ with $|v_{\mathrm{RF},m}|=1,\ m=1,...,N_{\mathrm{T}}$. Let $\tilde{\bm{A}}=\bm{A}_1(\bm{W}_{\mathrm{RF}}^H)$. Minimizing $\mathrm{PCRB}_{\theta}$ is equivalent to maximizing $\bm{v}_{\mathrm{RF}}^H\tilde{\bm{A}}\bm{v}_{\mathrm{RF}}=2\mathfrak{Re}\{v_{\mathrm{RF},m}\sum_{i\neq m}^{N_\mathrm{T}}v_{\mathrm{RF},i}^*[\tilde{\bm{A}}]_{i,m}\}+\sum_{i\neq m}^{N_{\mathrm{T}}}\sum_{j\neq i,j\neq m}^{N_{\mathrm{T}}}v_{\mathrm{RF},i}^*[\tilde{\bm{A}}]_{i,j}v_{\mathrm{RF},j}+\mathrm{tr}(\tilde{\bm{A}})$. Based on this, the optimal solution to each $v_{\mathrm{RF},m}$ with the other optimization variables being fixed can be shown to be given by
    \begin{numcases}
        {v_{\mathrm{RF},m}^{\star}\!\!=\!\!} 1, \!\!&\! if $\sum_{i\neq m}^{N_\mathrm{T}}v_{\mathrm{RF},i}[\tilde{\bm{A}}]_{i,m}^*\!\!=\!0$\!\!\!\!\!\!\!\!\!\!\nonumber\\
        \frac{\sum_{i\neq m}^{N_\mathrm{T}}v_{\mathrm{RF},i}[\tilde{\bm{A}}]_{i,m}^*}{|\sum_{i\neq m}^{N_\mathrm{T}}v_{\mathrm{RF},i}[\tilde{\bm{A}}]_{i,m}^*|}, \hspace{-1em}&\! otherwise.\!\!\!\!\! \label{vRF_solution}
    \end{numcases}

\subsubsection{Overall AO-based algorithm for (P0)}
To summarize, when $N_{\mathrm{RF},\mathrm{T}}=1$, the optimal transmit digital beamforming is given by ${R}_{\mathrm{BB}}^{\star}=\frac{P}{N_{\mathrm{T}}}$, and the AO-based algorithm iteratively obtains the optimal solution to each element in $\bm{V}_{\mathrm{RF}}$ (or $\bm{v}_{\mathrm{RF}}$) via (\ref{vRF_solution}) or each element in $\bm{W}_{\mathrm{RF}}^H$ (or $\bm{d}$) via (\ref{wRF_solution}) with the other variables being fixed at each time. When $N_{\mathrm{RF},\mathrm{T}}\geq2$, the AO-based algorithm iteratively optimizes $\bm{V}_{\mathrm{RF}}$ and $\bm{R}_{\mathrm{BB}}$ in a similar manner as Case II in Section \ref{subsubsectionCaseII} and each element in $\bm{W}_{\mathrm{RF}}^H$ (or $\bm{d}$) via (\ref{wRF_solution}). Note that since the optimal solution to each (block of) optimization variable is obtained and variables in different optimization blocks are not coupled in the constraints, the proposed algorithm is guaranteed to converge to a KKT point of (P0) \cite{solodov1998convergence}.\looseness=-1

Let $\mathcal{O}(\bar{\omega})$ denote the complexity of integrating a one-dimensional function, and $N_{\mathrm{out}}$ denote the number of outer iterations. The complexity  of the proposed algorithm for (P0) can be shown to be $\mathcal{O}(N_{\mathrm{out}}(N_{\mathrm{R}}^2\bar{\omega}+N_{\mathrm{R}}^2N_{\mathrm{T}}+N_{\mathrm{T}}^2\bar{\omega}+N_{\mathrm{T}}^2N_{\mathrm{R}}+N_{\mathrm{R}}N_{\mathrm{T}}N_{\mathrm{RF},\mathrm{R}}))$ for the case of $N_{\mathrm{RF},\mathrm{T}}=1$ and $\mathcal{O}(N_{\mathrm{out}}(N_{\mathrm{R}}^2\bar{\omega}+N_{\mathrm{R}}^2N_{\mathrm{T}}+N_{\mathrm{R}}N_{\mathrm{T}}N_{\mathrm{RF},\mathrm{T}}+N_{\mathrm{R}}N_{\mathrm{RF},\mathrm{T}}^2+N_{\mathrm{T}}^2\bar{\omega}+N_{\mathrm{T}}^2N_{\mathrm{R}}+N_{\mathrm{T}}N_{\mathrm{R}}N_{\mathrm{RF},{\mathrm{R}}}+N_{\mathrm{T}}^3))$ for the case of $N_{\mathrm{RF},\mathrm{T}}\geq2$. 

Notice from (\ref{wRF_solution}) and (\ref{vRF_solution}) that the optimal solution to each element in the transmit/receive analog beamforming matrix for the sensing function is critically dependent on the PDF of $\theta$. This demonstrates the important role of prior distribution information in the hybrid beamforming design, which will also be shown numerically in Section \ref{sec_numerical results}.

\vspace{-4mm}
\section{Hybrid Beamforming for MIMO ISAC}\label{sec_narrowband}
\vspace{-1mm}
In this section, we study the hybrid beamforming optimization for MIMO ISAC with both sensing and communication functions, which is more challenging than the sensing-only case in Section \ref{sec_sensing only}. Specifically, to minimize $\mathrm{PCRB}_{\theta}$, $\bm{V}_{\mathrm{RF}}$ and $\bm{R}_{\mathrm{BB}}$  need to be designed to ``align'' with the optimized $\bm{A}_1(\bm{W}_{\mathrm{RF}}^H)$ in which $\bm{W}_{\mathrm{RF}}^H$ needs to be optimized based on the steering vectors in $\bm{M}(\theta)$ and probability densities for different $\theta$'s. On the other hand, to maximize the communication rate, $\bm{V}_{\mathrm{RF}}$ and $\bm{R}_{\mathrm{BB}}$ should be designed to ``align'' with the MIMO communication channel $\bm{H}$. Due to the distinct objectives in sensing and communication, the hybrid beamforming design needs to strike an optimal balance between the two functions, thereby being highly non-trivial. 

\vspace{-5mm}
\subsection{Problem Formulation}
\vspace{-1mm}
We aim to jointly optimize the BS transmit hybrid beamforming in $\bm{V}_{\mathrm{RF}}$ and $\bm{R}_{\mathrm{BB}}$ and the BS receive analog beamforming in $\bm{W}_{\mathrm{RF}}^H$ to minimize the PCRB in sensing $\theta$, subject to a MIMO communication rate constraint specified by $\Bar{R}$ nps/Hz. The optimization problem is formulated as
		\begin{align}
			\mbox{(P1)} \underset{\substack{\bm{V}_{\mathrm{RF}},\bm{R}_{\mathrm{BB}}\succeq\bm{0},\\\bm{W}_{\mathrm{RF}}^H:(\ref{WRF_constraint})}}{\min}& \mathrm{PCRB}_{\theta} \label{P1_obj}\\
			\mbox{s.t.}\quad\quad & \log\bigg|\bm{I}_{N_{\mathrm{U}}}+\frac{\bm{H}\bm{V}_{\mathrm{RF}}\bm{R}_{\mathrm{BB}}\bm{V}_{\mathrm{RF}}^H\bm{H}^H}{\sigma_{\mathrm{C}}^2}\bigg|\geq \Bar{R} \label{P1_rate constraint}\\
			& \mathrm{tr}(\bm{V}_{\mathrm{RF}}\bm{R}_{\mathrm{BB}}\bm{V}_{\mathrm{RF}}^H)\leq P \label{P1_power constraint}\\
                & |[\bm{V}_{\mathrm{RF}}]_{m,n}|=1,\quad \forall m,n. \label{P1_unit-modulus constraint1}
		\end{align}
Note that (P1) is a non-convex problem due to the unit-modulus constraints in (\ref{WRF_constraint}) and (\ref{P1_unit-modulus constraint1}) as well as the coupling among $\bm{V}_{\mathrm{RF}}$, $\bm{R}_{\mathrm{BB}}$, and $\bm{W}_{\mathrm{RF}}^H$ via (\ref{P1_obj}) and (\ref{P1_rate constraint}). Due to the new PCRB expression in (\ref{P1_obj}), existing methods for dealing with the unit-modulus constraints (e.g., \cite{sohrabi2016hybrid, sohrabi2017hybrid, zhang2020capacity}) are not applicable.
To tackle this problem, we develop an AO-based algorithm to iteratively optimize the transmit digital beamforming in $\bm{R}_{\mathrm{BB}}$, the transmit analog beamforming matrix $\bm{V}_{\mathrm{RF}}$, and each element in the receive analog beamforming matrix $\bm{W}_{\mathrm{RF}}^H$ with the other variables being fixed at each time, as detailed below.

\vspace{-4.5mm}
\subsection{Proposed AO-based Solution to (P1)}
\vspace{-2mm}
\subsubsection{Optimization of $\bm{V}_{\mathrm{RF}}$ with given $\bm{R}_{\mathrm{BB}}$ and $\bm{W}_{\mathrm{RF}}^H$}\label{section_narrowband_VRF}
With given $\bm{R}_{\mathrm{BB}}$ and $\bm{W}_{\mathrm{RF}}^H$, (P1) is reduced to the following sub-problem for optimizing $\bm{V}_{\mathrm{RF}}$:
		\begin{align}
			\mbox{(P1-I)}\underset{\bm{V}_\mathrm{RF}:(\ref{P1_rate constraint})-(\ref{P1_unit-modulus constraint1})}{\max} \mathrm{tr}\big(\tilde{\bm{A}}\bm{V}_{\mathrm{RF}}\bm{R}_{\mathrm{BB}}\bm{V}_{\mathrm{RF}}^H\big), \label{P1I_obj}
		\end{align}
where $\tilde{\bm{A}}=\bm{A}_1(\bm{W}_{\mathrm{RF}}^H)$. Note that (P1-I) is still a non-convex problem due to the unit-modulus constraints in (\ref{P1_unit-modulus constraint1}). Furthermore, the left-hand side (LHS) of the rate constraint in (\ref{P1_rate constraint}) is not a concave function with respect to $\bm{V}_{\mathrm{RF}}$. Inspired by the \emph{WMMSE} method \cite{shi2011iteratively}, we propose to convert the rate constraint in (\ref{P1_rate constraint}) into a more tractable form as follows. 

We first introduce a decoding matrix $\bm{Q}^H\in \mathbb{C}^{N_{\mathrm{S}}\times N_{\mathrm{U}}}$. The decoded signal vector at the $l$-th interval is given by $\Hat{\bm{s}}_l=\bm{Q}^H\bm{y}_l^{\mathrm{C}}$. The MSE matrix for information decoding is given by \looseness=-1
     \begin{align}\label{MSE}
    	\bm{E}=&\mathbb{E}\Big[(\Hat{\bm{s}}_l-\bm{s}_l)(\Hat{\bm{s}}_l-\bm{s}_l)^H\Big]=\sigma_{\mathrm{C}}^2\bm{Q}^H\bm{Q} \\
    	&+(\bm{Q}^H\bm{H}\bm{V}_{\mathrm{RF}}\bm{V}_{\mathrm{BB}}-\bm{I}_{N_{\mathrm{S}}})(\bm{Q}^H\bm{H}\bm{V}_{\mathrm{RF}}\bm{V}_{\mathrm{BB}}-\bm{I}_{N_{\mathrm{S}}})^H. \nonumber
    \end{align}  
The communication rate can be expressed as
    \begin{equation}\label{reformulate_rate constraint}
		\xi(\bm{U}, \bm{Q}, \bm{V}_{\mathrm{RF}})\!=\!\log\big|\bm{U}\big|\!-\!\mathrm{tr}(\bm{U}\bm{E})\!+\!N_{\mathrm{S}},
    \end{equation}
where $\bm{U}\in \mathbb{C}^{N_{\mathrm{S}}\times N_{\mathrm{S}}}$ is an auxiliary matrix. 	
With given $\bm{V}_{\mathrm{RF}}$, the optimal $\bm{Q}$ for maximizing $\xi(\bm{U}, \bm{Q}, \bm{V}_{\mathrm{RF}})$ can be derived based on the first-order optimality condition as $\bm{Q}^{\star}=\bm{J}^{-1}\bm{H}\bm{V}_{\mathrm{RF}}\bm{V}_{\mathrm{BB}}$, with $\bm{J}=\sigma_{\mathrm{C}}^2\bm{I}_{N_{\mathrm{U}}}+\bm{H}\bm{V}_{\mathrm{RF}}\bm{V}_{\mathrm{BB}}\bm{V}_{\mathrm{BB}}^H\bm{V}_{\mathrm{RF}}^H\bm{H}^H$. With given $\bm{Q}$ and $\bm{V}_{\mathrm{RF}}$, the optimal $\bm{U}$ for maximizing $\xi(\bm{U},\bm{Q},\bm{V}_{\mathrm{RF}})$ is given by $\bm{U}^{\star}=\bm{E}^{-1}$. With the optimized $\bm{Q}$ and $\bm{U}$, $\xi(\bm{U}, \bm{Q}, \bm{V}_{\mathrm{RF}})$ is equal to the original communication rate expression in the LHS of (\ref{P1_rate constraint}). Based on this, we have the following proposition.
\begin{proposition}\label{proposition2}
    (P1-I) is equivalent to the following problem: 
		\begin{align}
			\mbox{(P1-I-eqv)}\underset{\bm{V}_\mathrm{RF}:(\ref{P1_power constraint}),(\ref{P1_unit-modulus constraint1}),\bm{Q},\bm{U}}{\max}\ & \mathrm{tr}\big(\tilde{\bm{A}}\bm{V}_{\mathrm{RF}}\bm{R}_{\mathrm{BB}}\bm{V}_{\mathrm{RF}}^H\big) \label{P1I'_obj}\\
			\mbox{s.t.}\qquad\quad\ & \xi(\bm{U}, \bm{Q}, \bm{V}_{\mathrm{RF}})\geq\Bar{R}.\label{P1I'_rate constraint}
		\end{align}
\end{proposition}
\begin{IEEEproof}
    For any feasible solution of (P1-I) denoted by $\bm{V}_{\mathrm{RF}}'$, we can find a solution of (P1-I-eqv) denoted as $(\bm{Q}^{\star}, \bm{U}^{\star}, \bm{V}_{\mathrm{RF}}')$ via $\bm{Q}^{\star}=\bm{J}^{-1}\bm{H}\bm{V}_{\mathrm{RF}}'\bm{V}_{\mathrm{BB}}$ and $\bm{U}^{\star}=\bm{E}^{-1}$. Since $\xi(\bm{Q}^{\star}, \bm{U}^{\star}, \bm{V}_{\mathrm{RF}}')=R\geq \Bar{R}$, $(\bm{Q}^{\star}, \bm{U}^{\star}, \bm{V}_{\mathrm{RF}}')$ is a feasible solution to (P1-I-eqv) which achieves the same objective value of (P1-I-eqv) as that of (P1-I) with $\bm{V}_{\mathrm{RF}}'$. For any feasible solution of (P1-I-eqv) denoted by $(\bm{Q}', \bm{U}', \bm{V}_{\mathrm{RF}}')$, since we have $\Bar{R}\leq\xi(\bm{Q}', \bm{U}', \bm{V}_{\mathrm{RF}}')\leq\xi(\bm{Q}^{\star}, \bm{U}^{\star}, \bm{V}_{\mathrm{RF}}')=R$ with $\bm{Q}^{\star}=\bm{J}^{-1}\bm{H}\bm{V}_{\mathrm{RF}}'\bm{V}_{\mathrm{BB}}$ and $\bm{U}^{\star}=\bm{E}^{-1}$, $\bm{V}_{\mathrm{RF}}'$ is a feasible solution to (P1-I) which achieves the same objective value as (P1-I-eqv) with $(\bm{Q}', \bm{U}', \bm{V}_{\mathrm{RF}}')$. Proposition \ref{proposition2} is thus proved. \looseness=-1
\end{IEEEproof}

Despite being non-convex, (P1-I-eqv) can be dealt with under the AO framework by iteratively optimizing $\bm{V}_{\mathrm{RF}}$, $\bm{Q}$, and $\bm{U}$, where the only remaining task is to optimize $\bm{V}_{\mathrm{RF}}$ with given $\bm{Q}$ and $\bm{U}$. By substituting $\bm{E}$ into (\ref{reformulate_rate constraint}), we have
    \begin{align}
		\xi(\bm{U}, \bm{Q}, \bm{V}_{\mathrm{RF}})=
		&\eta-\mathrm{tr}\big(\bm{V}_{\mathrm{BB}}^H\bm{V}_{\mathrm{RF}}^H\bm{B}_1\bm{V}_{\mathrm{RF}}\bm{V}_{\mathrm{BB}}\big)\nonumber\\
		&+\mathrm{tr}\big(\bm{B}_2\bm{V}_{\mathrm{RF}}\big)+\mathrm{tr}\big(\bm{B}_2^H\bm{V}_{\mathrm{RF}}^H\big),
    \end{align}
where $\bm{B}_1=\bm{H}^H\bm{Q}\bm{U}\bm{Q}^H\bm{H}$, $\bm{B}_2=\bm{V}_{\mathrm{BB}}\bm{U}\bm{Q}^H\bm{H}$, and  $\eta=\log|\bm{U}|+N_{\mathrm{S}}-\mathrm{tr}\big(\bm{U}\big)-\sigma_{\mathrm{C}}^2\mathrm{tr}\big(\bm{U}\bm{Q}^H\bm{Q}\big)$. Since $\mathrm{tr}(\bm{A}\bm{B}\bm{C}\bm{D})=(\mathrm{vec}(\bm{D}^T))^T(\bm{C}^T\otimes\bm{A})\mathrm{vec}(\bm{B})$, the problem of optimizing $\bm{V}_{\mathrm{RF}}$ is equivalently expressed as
		\begin{align}
			\mbox{(P1-I-eqv')}\, \underset{\bm{v}}{\max}\ & 
			\bm{v}^H(\bm{R}_{\mathrm{BB}}^T\otimes \tilde{\bm{A}})\bm{v}\label{P1Ieqvm_obj}\\
			\mbox{s.t.}\quad &\bm{v}^H(\bm{R}_{\mathrm{BB}}^T\otimes \bm{I}_{N_{\mathrm{T}}})\bm{v}\leq P \label{P1Ieqvm_power constraint} \\   
			&\bm{v}^H\!(\bm{R}_{\mathrm{BB}}^T\!\otimes\! \bm{B}_{1})\bm{v}\!\!-\!2\mathfrak{Re}(\bm{c}^T\bm{v})\!\leq\!\! -\Bar{R}\!+\!\eta \label{P1Ieqvm_rate constraint}\\
			& \bm{v}^H \bm{\Lambda}_m\bm{v}= 1, \, m=1,...,N_{\mathrm{T}}N_{\mathrm{RF},\mathrm{T}},\label{P1Ieqvm_unit_modulus_constraint}
		\end{align}
where $\bm{v}=\mathrm{vec}(\bm{V}_{\mathrm{RF}})$; $\bm{c}=\mathrm{vec}(\bm{B}_2^T)$; $\bm{\Lambda}_m\in \mathbb{R}^{N_{\mathrm{T}}N_{\mathrm{RF},\mathrm{T}}\times N_{\mathrm{T}}N_{\mathrm{RF},\mathrm{T}} }$ is a matrix whose $m$-th diagonal element is $1$ and all the remaining elements are $0$. 
	
(P1-I-eqv') is a non-convex quadratically constrained quadratic program (QCQP) since $\bm{R}_{\mathrm{BB}}^T\otimes \tilde{\bm{A}}$ is a positive semi-definite matrix and $\bm{\Lambda}_m \neq \bm{0}$. To tackle this problem, we propose an \emph{FPP-SCA} based approach \cite{mehanna2014feasible}. 
Specifically, we first note that each $m$-th constraint in (\ref{P1Ieqvm_unit_modulus_constraint}) can be equivalently written as $\bm{v}^H \bm{\Lambda}_m\bm{v}\leq 1$ and $\bm{v}^H \bm{\Lambda}_m\bm{v}\geq 1$, where the latter is non-convex. Then, we apply the SCA technique to replace the latter and the objective function in (\ref{P1Ieqvm_obj}) with their first-order Taylor approximations, respectively. Finally, we introduce slack variables $r$, $\bm{p}\in\mathbb{C}^{N_{\mathrm{T}}N_{\mathrm{RF},\mathrm{T}}\times 1}$, and $\bm{w}\in\mathbb{C}^{N_{\mathrm{T}}N_{\mathrm{RF},\mathrm{T}}\times 1}$ to further expand the search space in each iteration to enable higher flexibility in pursuing a high-quality feasible point.\footnote{Note that this is the major difference between FPP-SCA and SCA, and is particularly suitable for the case where a feasible initial point cannot be found for SCA. In our case, a feasible solution of $\bm{v}$ can be easily found thanks to the unit-modulus constraint, while the FPP-SCA based approach can still achieve enhanced performance due to increased flexibility brought by slack variables. \looseness=-1} By further introducing $t$ as an auxiliary variable and $\epsilon\gg 1$ as the penalty factor, the following problem will be iteratively solved under the FPP-SCA framework with a given local point $\bar{\bm{v}}$:
		\begin{align}
			\!\!\!\mbox{(P1-I-F)}\!\!\!\!\!\!\!\!\! \underset{\substack{\bm{v}:(\ref{P1Ieqvm_power constraint}),(\ref{P1Ieqvm_rate constraint}), t,\\ r\geq 0,\bm{p}\succeq\bm{0}, \bm{w}\succeq\bm{0}}}{\max}\!\!\!\!\!\!\!\!& 
			t-\epsilon r-\epsilon \|\bm{p}\|_1-\epsilon\|\bm{w}\|_1 \label{P1If_obj}\\
			\mbox{s.t.}\quad & 2\mathfrak{Re}\{\bar{\bm{v}}^H(\bm{R}_{\mathrm{BB}}^T\otimes \tilde{\bm{A}})\bm{v} \}\!-\!\bar{\bm{v}}^H(\bm{R}_{\mathrm{BB}}^T\otimes \tilde{\bm{A}})\bar{\bm{v}}\notag\\
			&\qquad\qquad\geq t-r\\
			& \bm{v}^H \bm{\Lambda}_m\bm{v}\leq 1+p_m, \quad \forall m\\
			\!\!& 2\mathfrak{Re}\big\{\bar{\bm{v}}^H\bm{\Lambda}_m\bm{v}\big\}\!-\!\bar{\bm{v}}^H\bm{\Lambda}_m\bar{\bm{v}}\!\geq\! 1\!-\!w_m, \, \forall m.
		\end{align}
(P1-I-F) is a convex optimization problem, which can be efficiently solved via the interior-point method \cite{CVX} or existing software such as CVX \cite{cvxtool}. By iteratively solving (P1-I-F) with $\bar{\bm{v}}$  updated as the optimal solution to (P1-I-F) in the previous iteration, the FPP-SCA approach is guaranteed to converge monotonically. Moreover, it is guaranteed to converge to a KKT point of (P1-I-eqv') as long as the slack variables are converged to zero \cite{mehanna2014feasible}.

\subsubsection{Optimal solution to $\bm{R}_{\mathrm{BB}}$}\label{section_narrowband_RBB}   
With given $\bm{V}_{\mathrm{RF}}$ and $\bm{W}_{\mathrm{RF}}^H$, the transmit digital beamforming in $\bm{R}_{\mathrm{BB}}$ can be optimized via solving the following problem:
		\begin{align}
			\mbox{(P1-II)}\underset{\bm{R}_{\mathrm{BB}}\succeq\bm{0}:(\ref{P1_rate constraint}),(\ref{P1_power constraint})}{\max}\ & \mathrm{tr}\big(\tilde{\bm{A}}\bm{V}_{\mathrm{RF}}\bm{R}_{\mathrm{BB}}\bm{V}_{\mathrm{RF}}^H\big). \label{P1II_obj}
		\end{align}
Note that (P1-II) is a convex optimization problem. The optimal solution to (P1-II) denoted by $\bm{R}_{\mathrm{BB}}^{\star}$ can be derived via the Lagrange duality method as
    \begin{align} \label{RBB_solution}
    	\bm{R}_{\mathrm{BB}}^\star \!=\! \left\{
    	\begin{array}{ll}
    		\!P\bm{V}_{\mathrm{RF}}^{\dagger}\bm{q}_1\bm{q}_1^H\bm{V}_{\mathrm{RF}}^{\dagger H},  & \bar{R}\leq R_{\mathrm{S}}     \\
    		\!\tilde{{\bm{Q}}}^{ -\frac{1}{2}}\tilde{\bm{V}}\tilde{\bm{\Lambda}}\tilde{\bm{ V}}^H\tilde{{\bm{Q}}}^{ -\frac{1}{2}},&\bar{R}\geq R_{\mathrm{S}}.
    	\end{array}
    	\right.
    \end{align}
Specifically, $\bm{q}_1$ denotes the eigenvector corresponding to the largest eigenvalue of $\tilde{\bm{A}}$ and  $R_{\mathrm{S}}=\log\left|\bm{I}_{N_{\mathrm{U}}}+ \frac{P\bm{H}\bm{q}_1\bm{q}_1^H\bm{H}^H}{\sigma_{\mathrm{C}}^2}\right|$;  $\tilde{\bm{Q}}=\bm{V}_{\mathrm{RF}}^H(\frac{\mu^\star}{\beta^\star}\bm{I}_{N_{\mathrm{T}}}-\frac{1}{\beta^\star}\tilde{\bm{A}})\bm{V}_{\mathrm{RF}}$ where $\beta^{\star}$ and $\mu^{\star}$ denote the optimal dual variables associated with the constraints in (\ref{P1_rate constraint}) and (\ref{P1_power constraint}), respectively; 
$\tilde{\bm{H}}\tilde{\bm{Q}}^{-\frac{1}{2}}= \tilde{\bm{U}}\tilde{\bm{\Gamma}}^{\frac{1}{2}}\tilde{\bm{V}}^H$ denotes the (reduced) SVD of $\tilde{\bm{H}}\tilde{\bm{Q}}^{-\frac{1}{2}}$, where $\tilde{\bm{H}}=\bm{H}\bm{V}_{\mathrm{RF}}$, $\tilde{\bm{\Gamma}}=\mathrm{diag}\{\tilde{h}_1,...,\tilde{h}_{T}\}$, $T=\mathrm{rank}(\tilde{\bm{H}})$, $\tilde{\bm{U}} \in \mathbb{C}^{N_{\mathrm{U}}\times T}$ and $\tilde{\bm{V}}\in \mathbb{C}^{N_{\mathrm{RF},\mathrm{T}}\times T}$ with $\tilde{\bm{U}}\tilde{\bm{U}}^H= \bm{I}_{N_{\mathrm{U}}} $ and $\tilde{\bm{V}}\tilde{\bm{V}}^H= \bm{I}_{N_{\mathrm{RF},\mathrm{T}}}$, and $\tilde{\bm{ \Lambda}}=\mathrm{diag}\{\tilde{v}_1,...,\tilde{v}_{T}\}$ with $\tilde{v}_i=(1/\ln2-\sigma_{\mathrm{C}}^2/\tilde{h}_i)^{+},\ i=1,2,...,T$.\footnote{This problem is similar to (P4) in \cite{xu2024mimo}, where more details can be found.} \looseness=-1

\subsubsection{Optimal solution to each element in $\bm{W}_{\mathrm{RF}}^H$}
With given $\bm{R}_{\mathrm{BB}}$ and $\bm{V}_{\mathrm{RF}}$, the subproblem for  optimizing $\bm{W}_{\mathrm{RF}}^H$ has the same structure as that of the sensing-only case. Thus, the optimal solution to each element of $\bm{W}_{\mathrm{RF}}^H$ can be obtained in a similar manner as that in Section \ref{sec_Optimize_WRF} via (\ref{wRF_solution}). \looseness=-1

\subsubsection{Overall AO-based algorithm for (P1)}
To summarize, the overall AO-based algorithm for (P1) iteratively optimizes the transmit analog beamforming by leveraging WMMSE transformation and the FPP-SCA technique, the transmit digital beamforming via (\ref{RBB_solution}), and each element in the receive analog beamforming matrix $\bm{W}_{\mathrm{RF}}^H$ via (\ref{wRF_solution}). An initial point can be obtained by randomly generating multiple independent  realizations of $\bm{V}_{\mathrm{RF}}$ and $\bm{W}_{\mathrm{RF}}^H$ under a uniform distribution for the phase of each unit-modulus element therein, obtaining the optimal $\bm{R}_{\mathrm{BB}}$ via solving (P1-II) for each realization, and finally selecting the best one that is both feasible for (P1) and achieves the minimum objective value. Note that the objective function of (P1) is guaranteed to be decreased after each update in the AO-based algorithm, thus the AO-based algorithm is guaranteed to converge monotonically. 
  
Let $N_{\mathrm{in}}$ denote the number of iterations in FPP-SCA, $N_{\mathrm{V}}$ denote the number of iterations for optimizing $\bm{Q}$, $\bm{U}$, and $\bm{V}_{\mathrm{RF}}$ under WMMSE, $N_{\mathrm{LD}}$ denote the number of iterations in the Lagrange duality method for solving (P1-II), and $N_{\mathrm{out}}$ denote the number of outer iterations. The worst-case complexity for the proposed algorithm can be shown to be
$\mathcal{O}(N_{\mathrm{out}}(N_{\mathrm{R}}^2\bar{\omega}+N_{\mathrm{T}}^2\bar{\omega}+N_{\mathrm{R}}N_{\mathrm{T}}^2+N_{\mathrm{R}}^2N_{\mathrm{T}}+N_{\mathrm{R}}N_{\mathrm{T}}N_{\mathrm{RF},\mathrm{R}}+N_{\mathrm{R}}N_{\mathrm{T}}N_{\mathrm{RF},\mathrm{T}}+N_{\mathrm{R}}N_{\mathrm{RF},\mathrm{T}}^2+N_{\mathrm{LD}}(N_{\mathrm{RF},\mathrm{T}}^3+N_{\mathrm{T}}^2N_{\mathrm{RF},\mathrm{T}}+N_{\mathrm{T}}N_{\mathrm{RF},\mathrm{T}}^2+N_{\mathrm{U}}N_{\mathrm{RF},\mathrm{T}}^2+N_{\mathrm
U}N_{\mathrm{RF},\mathrm{T}}\min(N_{\mathrm
U},N_{\mathrm{RF},\mathrm{T}}))+N_{\mathrm{V}}(N_{\mathrm{in}}(N_{\mathrm{T}}N_{\mathrm{RF},\mathrm{T}})^{3.5}+N_{\mathrm{U}}^3+N_{\mathrm{RF},\mathrm{T}}^2N_{\mathrm{U}}+N_{\mathrm{U}}N_{\mathrm{T}}^2+N_{\mathrm{U}}^2N_{\mathrm{T}}+N_{\mathrm{U}}N_{\mathrm{T}}N_{\mathrm{RF},\mathrm{T}})))$. \looseness=-1

\vspace{-3mm}
\section{Hybrid Beamforming for MIMO-OFDM ISAC}\label{sec_OFDM}
\vspace{-1mm}

In this section, we extend our results for narrowband systems to a \emph{wideband MIMO-OFDM ISAC} system under frequency-selective communication channel. In this case, an \emph{individual} digital beamforming matrix can be designed to cater to the communication channel at each OFDM sub-carrier, while a \emph{common} analog beamforming matrix needs to be designed for all sub-carriers. This thus brings extra design challenges as the frequency non-selective transmit analog beamforming needs to cater to the frequency-selective communication channels over multiple sub-carriers and the sensing function at the same time, especially under the challenging setup studied in this paper with hybrid array at the BS receiver for sensing.  

\vspace{-3mm}
\subsection{MIMO-OFDM ISAC System Model under Hybrid Arrays}
\vspace{-1mm}
We consider a wideband MIMO-OFDM ISAC system where the BS transmits signals over $K>1$ sub-carriers.
We focus on $L$ symbol intervals over which both the communication channel and the sensing target remain static. Let $N_{\mathrm{S},k}$ denote the number of data streams transmitted over the $k$-th sub-carrier, and $\bm{s}_{k,l}\in\mathbb{C}^{N_{\mathrm{S},k}\times 1}$ denote the data symbol vector at the $k$-th sub-carrier during the $l$-th symbol interval, with $\bm{s}_{k,l}\sim\mathcal{CN}(\bm{0}, \bm{I}_{N_{\mathrm{S},k}})$. The collection of the data symbols over $K$ sub-carriers and $L$ symbol intervals is given by $\bar{\bm{S}}=[\bm{S}_1^T,...,\bm{S}_K^T]^T\in\mathbb{C}^{N_\mathrm{S}\times L}$,
 where $\bm{S}_k=[\bm{s}_{k,1},...,\bm{s}_{k,L}]\in\mathbb{C}^{N_{\mathrm{S},k}\times L}$. $\bm{s}_{k,l}$ is first processed using a digital beamforming matrix $\bm{V}_{\mathrm{BB},k}\in\mathbb{C}^{N_{\mathrm{RF},\mathrm{T}}\times N_{\mathrm{S},k}}$  in the frequency domain, and then transformed into the time domain using an $N_{\mathrm{RF},\mathrm{T}}$ $K$-point inverse fast Fourier transform (IFFT).  After that, the signal is then processed with an analog beamforming matrix $\bm{V}_{\mathrm{RF}}\in\mathbb{C}^{N_\mathrm{T}\times N_{\mathrm{RF},\mathrm{T}}}$ in the time domain, which is common for all sub-carriers \cite{sohrabi2017hybrid}. The final transmit signal at sub-carrier $k$ in each $l$-th symbol interval is given by
\begin{equation}\label{OFDM_xkl}
    \bm{x}_{k,l}=\bm{V}_{\mathrm{RF}}\bm{V}_{\mathrm{BB},k}\bm{s}_{k,l},\quad l=1,...,L.
\end{equation}
The transmit covariance matrix over the $k$-th sub-carrier is \looseness=-1
\begin{align}
        \tilde{\bm{R}}_{X,k}=\mathbb{E}[\bm{x}_{k,l}\bm{x}_{k,l}^H]=\bm{V}_{\mathrm{RF}}\bm{R}_{\mathrm{BB},k}\bm{V}_{\mathrm{RF}}^H,
\end{align}
where $\bm{R}_{\mathrm{BB},k}=\bm{V}_{\mathrm{BB},k}\bm{V}_{\mathrm{BB},k}^H$.
Let $\bm{H}_k\in\mathbb{C}^{N_{\mathrm{U}}\times N_{\mathrm{T}}}$ denote the channel matrix at the $k$-th sub-carrier. The received signal at the $k$-th sub-carrier in each $l$-th symbol interval is given by 
\begin{equation}
    \bm{y}_{k,l}^{\mathrm{C}}=\bm{H}_k\bm{V}_{\mathrm{RF}}\bm{V}_{\mathrm{BB},k}\bm{s}_{k,l}+\bm{n}^{\mathrm{C}}_{k,l},\quad l=1,...,L,
\end{equation}
where $\bm{n}^{\mathrm{C}}_{k,l}\sim\mathcal{CN}(\bm{0}, \bar{\sigma}_{\mathrm{C}}^2\bm{I}_{N_{\mathrm{U}}})$ denotes the CSCG noise vector at the user receiver with $\bar{\sigma}_{\mathrm{C}}^2$ denoting the average noise power at each sub-carrier.
The achievable rate of sub-carrier $k$ is given by
\begin{equation}\label{OFDM_rate_k}
    R_k^{\mathrm{OFDM}}=\frac{1}{K}\log\bigg|\bm{I}_{N_{\mathrm{U}}}+\frac{\bm{H}_k\bm{V}_{\mathrm{RF}}\bm{R}_{\mathrm{BB},k}\bm{V}_{\mathrm{RF}}^H\bm{H}_k^H}{\bar{\sigma}_{\mathrm{C}}^2}\bigg|
\end{equation}
in nps/Hz. The achievable rate over the $K$ sub-carriers is then given by $R^{\mathrm{OFDM}}=\sum_{k=1}^K R_k^{\mathrm{OFDM}}$
in nps/Hz.\footnote{We neglect the constant factor for cyclic prefix for simplicity.}

By sending downlink signals, the BS can also estimate $\theta$ based on the received echo signals and the prior distribution information. With a common receive analog beamforming matrix $\bm{W}_{\mathrm{RF}}^H$ for all sub-carriers, the effective received signal vector at the $k$-th sub-carrier in each $l$-th symbol interval is given by \looseness=-1
\begin{align}\label{OFDM_y_kl}
        \tilde{\bm{y}}_{k,l}^{\mathrm{S}}=&\bm{W}_{\mathrm{RF}}^H\alpha\bm{b}(\theta)\bm{a}^H(\theta)
    \bm{V}_{\mathrm{RF}}\bm{V}_{\mathrm{BB},k}\bm{s}_{k,l}+\bm{W}_{\mathrm{RF}}^H\bm{n}_{k,l}^{\mathrm{S}},
\end{align}
where we assume the complex reflection coefficient is $\alpha$ for all sub-carriers.\footnote{Note that this assumption is valid for practical OFDM systems where the operating frequency is much larger than the system bandwidth, thus the signal wavelengths in different sub-carriers are approximately the same \cite{sohrabi2017hybrid}.} 
The equivalent noise vector is given by $\bm{z}_{k,l}=\bm{W}_{\mathrm{RF}}^H\bm{n}_{k,l}^{\mathrm{S}}\sim\mathcal{CN}(\bm{0}, \bar{\sigma}_{\mathrm{S}}^2\bm{W}_{\mathrm{RF}}^H\bm{W}_{\mathrm{RF}})$ with $\bar{\sigma}_{\mathrm{S}}^2$ denoting the average noise power at each sub-carrier.
 Let $\bm{V}_{\mathrm{BB},\mathrm{blk}}=\mathrm{blkdiag}(\bm{V}_{\mathrm{BB},1},...,\bm{V}_{\mathrm{BB},K})\in\mathbb{C}^{N_{\mathrm{RF},\mathrm{T}}K\times N_{\mathrm{S}}}$ denote the collection of transmit digital beamforming matrices, and $\bm{V}_{\mathrm{RF},\mathrm{blk}}=\mathrm{blkdiag}(\bm{V}_{\mathrm{RF}},...\bm{V}_{\mathrm{RF}})\in\mathbb{C}^{N_\mathrm{T}K\times N_{\mathrm{RF},\mathrm{T}}K}$ denote the collection of the (common) transmit analog beamforming matrices for all sub-carriers. The effective received signal matrix available for sensing can be expressed as
\begin{align}\label{OFDM_Y1}
    \tilde{\bm{Y}}^{\mathrm{S}}_{\mathrm{OFDM}}=&\alpha\mathrm{blkdiag}(\bm{W}_{\mathrm{RF}}^H\bm{b}(\theta)\bm{a}^H(\theta),...,\bm{W}_{\mathrm{RF}}^H\bm{b}(\theta)\bm{a}^H(\theta))\nonumber\\
    &\times \bm{V}_{\mathrm{RF},\mathrm{blk}}\bm{V}_{\mathrm{BB},\mathrm{blk}}\bar{\bm{S}}+\bar{\bm{Z}},
\end{align}
where $\bar{\bm{Z}}=\mathrm{blkdiag}(\bm{W}_{\mathrm{RF}}^H,...,\bm{W}_{\mathrm{RF}}^H)\bar{\bm{N}}$, with $\bar{\bm{N}}=[\bm{N}_1^T,...,\bm{N}_K^T]^T\in\mathbb{C}^{N_{\mathrm{R}}K\times L}$, $\bm{N}_{k}=[\bm{n}_{k,1}^{\mathrm{S}},...,\bm{n}_{k,L}^{\mathrm{S}}]$.

\vspace{-3mm}
\subsection{Characterization of Sensing Performance for MIMO-OFDM ISAC under Hybrid Arrays Exploiting Prior Information}
\vspace{-1mm}

In this subsection, we aim to characterize the sensing performance for the MIMO-OFDM ISAC system. To this end, we first vectorize (\ref{OFDM_Y1}), which yields $\bar{\bm{y}}=\mathrm{vec}(\tilde{\bm{Y}}^{\mathrm{S}}_{\mathrm{OFDM}})=\bar{\bm{u}}+\bar{\bm{z}}\sim\mathcal{CN}(\bar{\bm{u}},\bar{\bm{C}}_y)$, with 
\begin{align}\label{noise_free_signal_OFDM}
    \bar{\bm{u}}=&\alpha\mathrm{vec}(\mathrm{blkdiag}(\bm{W}_{\mathrm{RF}}^H\bm{b}(\theta)\bm{a}^H(\theta),..., \nonumber\\
    &\bm{W}_{\mathrm{RF}}^H\bm{b}(\theta)\bm{a}^H(\theta))\bm{V}_{\mathrm{RF},\mathrm{blk}}\bm{V}_{\mathrm{BB},\mathrm{blk}}\bar{\bm{S}}),
\end{align}
$\bar{\bm{z}}=\mathrm{vec}(\bar{\bm{Z}})$, and $\bar{\bm{C}}_y=\frac{N_{\mathrm{R}}}{N_{\mathrm{RF},\mathrm{R}}}\bar{\sigma}_{\mathrm{S}}^2\bm{I}_{N_{\mathrm{RF},\mathrm{R}}KL}$. Note that the noise-free received signal in (\ref{noise_free_signal_OFDM}) contains both $\theta$ and $\alpha$ as unknown parameters, which are thus jointly sensed based on the observations in $\bar{\bm{y}}$ and the prior distribution information. In this case, the PFIM from the observation can be shown to be given by $\bm{F}_{\mathrm{O}}^{\mathrm{OFDM}}=\begin{bmatrix}
            \bar{J}_{\theta\theta} & \bar{\bm{J}}_{\theta\alpha} \\
            \bar{\bm{J}}_{\theta\alpha}^{H} & \bar{\bm{J}}_{\alpha\alpha}
        \end{bmatrix}$. Each block in $\bm{F}_{\mathrm{O}}^{\mathrm{OFDM}}$ can be derived as \looseness=-1
    \begin{flalign}
			\bar{J}_{\theta\theta}
            =&\frac{2N_{\mathrm{RF},\mathrm{R}}L\gamma}{N_{\mathrm{R}}\bar{\sigma}_{\mathrm{S}}^2}\sum_{k=1}^K\mathrm{tr}(\bm{A}_1(\bm{W}_{\mathrm{RF}}^H)\bm{V}_{\mathrm{RF}}\bm{R}_{\mathrm{BB},k}\bm{V}_{\mathrm{RF}}^H), \label{OFDM_J_thetatheta}\\
            \bar{\bm{J}}_{\theta\alpha}
            =&\frac{2N_{\mathrm{RF},\mathrm{R}}L}{N_{\mathrm{R}}\bar{\sigma}_{\mathrm{S}}^2}\mathbb{E}_{\zeta}\Big[
            \mathfrak{Re}\Big\{\!\alpha^*\nonumber\\
            &\times\!\!\!\sum_{k=1}^K\!\mathrm{tr}(\Dot{\bm{M}}^H\!(\theta)\bm{W}_{\mathrm{RF}}\bm{W}_{\mathrm{RF}}^H\bm{M}\!(\theta)\bm{V}_{\mathrm{RF}}\bm{R}_{\mathrm{BB},k}\bm{V}_{\mathrm{RF}}^H)[1,j]\!\Big\}
            \!\Big]\notag\\
            =&\bm{0},\label{OFDM_J_thetaalpha}\\
            \!\!\bar{\bm{J}}_{\alpha\alpha}
            =&\frac{2N_{\mathrm{RF},\mathrm{R}}L}{N_{\mathrm{R}}\bar{\sigma}_{\mathrm{S}}^2}\sum_{k=1}^K\mathrm{tr}(\bm{A}_2(\bm{W}_{\mathrm{RF}}^H)\bm{V}_{\mathrm{RF}}\bm{R}_{\mathrm{BB},k}\bm{V}_{\mathrm{RF}}^H)\bm{I}_2,\label{OFDM_J_alphaalpha}
        \end{flalign}
where $\gamma\triangleq\iint(\alpha_{\mathrm{R}}^2+\alpha_{\mathrm{I}}^2)p_{\alpha_{\mathrm{R}},\alpha_{\mathrm{I}}}(\alpha_{\mathrm{R}},\alpha_{\mathrm{I}})d\alpha_{\mathrm{R}}d\alpha_{\mathrm{I}}$, $\bm{A}_1(\bm{W}_{\mathrm{RF}}^H)\!=\!\int\Dot{\bm{M}}^H(\theta)\bm{W}_{\mathrm{RF}}\bm{W}_{\mathrm{RF}}^H\Dot{\bm{M}}(\theta)p_{\Theta}(\theta)d\theta\succeq\bm{0}$, and $\bm{A}_2(\bm{W}_{\mathrm{RF}}^H)\!=\!\int\bm{M}^H(\theta)\bm{W}_{\mathrm{RF}}\bm{W}_{\mathrm{RF}}^H\bm{M}(\theta)p_{\Theta}(\theta)d\theta\succeq\bm{0}$. The overall PFIM is thus given by $\bm{F}^{\mathrm{OFDM}}=\bm{F}_{\mathrm{O}}^{\mathrm{OFDM}}+\bm{F}_{\mathrm{P}}$.
The PCRB for the MSE in estimating the desired parameter $\theta$ under MIMO-OFDM ISAC is thus given by
    \begin{align}\label{PCRB_theta_OFDM}
        &\mathrm{PCRB}_{\theta}^{\mathrm{OFDM}}=[{\bm{F}^{\mathrm{OFDM}}}^{-1}]_{1,1}=1\bigg/\bigg(\mathbb{E}_\theta\Big[\Big(\frac{\partial\ln(p_{\Theta}(\theta))}{\partial\theta}\Big)^2\Big]\nonumber\\
        &+\frac{2N_{\mathrm{RF},\mathrm{R}}L\gamma}{N_{\mathrm{R}}\bar{\sigma}_{\mathrm{S}}^2}\sum_{k=1}^K\mathrm{tr}(\bm{A}_1(\bm{W}_{\mathrm{RF}}^H)\bm{V}_{\mathrm{RF}}\bm{R}_{\mathrm{BB},k}\bm{V}_{\mathrm{RF}}^H)
        \bigg).
    \end{align}

\vspace{-4mm}
\subsection{Problem Formulation}
\vspace{-1mm}
In this subsection, we formulate the hybrid beamforming optimization problem for MIMO-OFDM ISAC as follows:
		\begin{align}
			\!\!\!\!\mbox{(P2)}\!\!\!\!\!\!\!\! \underset{\substack{\bm{V}_{\mathrm{RF}},\bm{W}_{\mathrm{RF}}^H:(\ref{WRF_constraint})\\\{\bm{R}_{\mathrm{BB},k}\succeq\bm{0}\}_{k=1}^K }}{\min}\!\!\!\!\!\!\!\!& \mathrm{PCRB}_{\theta}^{\mathrm{OFDM}} \label{P2_obj}\\
			\mbox{s.t.}\quad & 
               \frac{1}{K}\!\!\sum_{k=1}^K\log\bigg|\bm{I}_{N_{\mathrm{U}}} 
               \!\!+\!\frac{\bm{H}_k\bm{V}_{\mathrm{RF}}\bm{R}_{\mathrm{BB},k}\bm{V}_{\mathrm{RF}}^H\bm{H}_k^H}{\bar{\sigma}_{\mathrm{C}}^2}\bigg|\!\!\geq\! \Bar{R}
                \label{P2_rate constraint}\\
			& \sum_{k=1}^K\mathrm{tr}(\bm{V}_{\mathrm{RF}}\bm{R}_{\mathrm{BB},k}\bm{V}_{\mathrm{RF}}^H)\leq P \label{P2_power constraint}\\
                & |[\bm{V}_{\mathrm{RF}}]_{m,n}|=1,\quad \forall m,n. \label{P2_unit-modulus constraint1}
		\end{align}
Note that (P2) is a non-convex problem due to the unit-modulus constraints in (\ref{WRF_constraint}) and (\ref{P2_unit-modulus constraint1}). Moreover, besides the coupling among the transmit/receive analog beamforming and transmit digital beamforming, the transmit digital beamforming matrices for different sub-carriers are coupled in the PCRB expression, the rate expression, and the transmit power constraint, which makes (P2) more challenging than its narrowband counterpart.

\vspace{-3mm}
\subsection{Proposed AO-based Solution to (P2)}
\vspace{-1mm}

In this subsection, we devise an AO-based algorithm for (P2), by iteratively optimizing the transmit analog beamforming matrix $\bm{V}_{\mathrm{RF}}$, the transmit digital beamforming matrices in $\{\bm{R}_{\mathrm{BB},k}\}_{k=1}^K$, or each element in the receive beamforming matrix $\bm{W}_{\mathrm{RF}}^H$ with the other variables being fixed at each time.

\subsubsection{Optimization of $\bm{V}_{\mathrm{RF}}$ with given $\{\bm{R}_{\mathrm{BB},k}\}_{k=1}^K$ and $\bm{W}_{\mathrm{RF}}^H$}
With given $\{\bm{R}_{\mathrm{BB},k}\}_{k=1}^K$ and $\bm{W}_{\mathrm{RF}}^H$, (P2) is reduced to the following sub-problem for optimizing $\bm{V}_{\mathrm{RF}}$:
		\begin{align}
			\!\!\!\mbox{(P2-I)}\,\underset{\bm{V}_{\mathrm{RF}}:(\ref{P2_rate constraint})-(\ref{P2_unit-modulus constraint1})}{\max}& \sum_{k=1}^K\mathrm{tr}(\tilde{\bm{A}}\bm{V}_{\mathrm{RF}}\bm{R}_{\mathrm{BB},k}\bm{V}_{\mathrm{RF}}^H), \label{P2I_obj}
		\end{align}
where $\tilde{\bm{A}}=\bm{A}_1(\bm{W}_{\mathrm{RF}}^H)$. We leverage the \emph{WMMSE} method to handle the non-concave rate constraint in the LHS of (\ref{P2_rate constraint}).
By introducing a decoding matrix $\bm{Q}_k^H\in \mathbb{C}^{N_{\mathrm{S},k}\times N_{\mathrm{U}}}$ for each $k$-th sub-carrier, the MSE matrix for information decoding at the $k$-th sub-carrier is given by \looseness=-1
\begin{align}
        \!\!\!\bm{E}_k\!\!&=\mathbb{E}[(\Hat{\bm{s}}_{k,l}-\bm{s}_{k,l})(\Hat{\bm{s}}_{k,l}-\bm{s}_{k,l})^H]
        =\bar{\sigma}_{\mathrm{C}}^2\bm{Q}_k^H\bm{Q}_k\nonumber\\
        &\!\!\!\!\!+\!(\!\bm{Q}_k^H\!\bm{H}_k\!\bm{V}_{\mathrm{RF}}\!\bm{V}_{\mathrm{BB},k}\!\!-\!\!\bm{I}_{N_\mathrm{S},k})(\bm{Q}_k^H\!\bm{H}_k\!\bm{V}_{\mathrm{RF}}\!\bm{V}_{\mathrm{BB},k}\!\!-\!\!\bm{I}_{N_\mathrm{S},k})^H\!\!\!. 
\end{align}
Then, $KR_k^{\mathrm{OFDM}}$ can be expressed as
\begin{equation}\label{OFDM_eqv_rate1}
        \xi_k(\bm{U}_k, \bm{Q}_k,\bm{V}_{\mathrm{RF}})=\log|\bm{U}_k|-\mathrm{tr}(\bm{U}_k\bm{E}_k)+N_{\mathrm{S},k},
\end{equation}
where $\bm{U}_k\in \mathbb{C}^{N_{\mathrm{S},k}\times N_{\mathrm{S},k}}$ is an auxiliary matrix. With $\bm{Q}_k$ optimized as $\bm{Q}_k^{\star}=\bm{J}_k^{-1}\bm{H}_k\bm{V}_{\mathrm{RF}}\bm{V}_{\mathrm{BB},k}$ where $\bm{J}_k=\bar{\sigma}_{\mathrm{C}}^2\bm{I}_{N_{\mathrm{U}}}+\bm{H}_k\bm{V}_{\mathrm{RF}}\bm{V}_{\mathrm{BB},k}\bm{V}_{\mathrm{BB},k}^H\bm{V}_{\mathrm{RF}}^H\bm{H}_k^H$ and $\bm{U}_k$ optimized as $\bm{U}_k^{\star}=\bm{E}_k^{-1}$, $\frac{1}{K}\xi_k(\bm{U}_k, \bm{Q}_k,\bm{V}_{\mathrm{RF}})$ is equal to $R_k^{\mathrm{OFDM}}$ in (\ref{OFDM_rate_k}). Problem (P2-I) is thus equivalent to the following problem:
		\begin{align}
			\!\!\!\!\mbox{(P2-I-eqv)}\!\!\!\!\underset{\substack{\bm{V}_{\mathrm{RF}}:(\ref{P2_power constraint}),(\ref{P2_unit-modulus constraint1}),\\ \{\bm{Q}_k\}_{k=1}^{K},\{\bm{U}_k\}_{k=1}^{K}}}{\max}& \sum_{k=1}^K\mathrm{tr}(\tilde{\bm{A}}\bm{V}_{\mathrm{RF}}\bm{R}_{\mathrm{BB},k}\bm{V}_{\mathrm{RF}}^H) \label{P2Ieq_obj}\\
			\mbox{s.t.}\qquad\quad & \frac{1}{K}\sum_{k=1}^K\xi_k(\bm{U}_k, \bm{Q}_k,\bm{V}_{\mathrm{RF}})\geq \Bar{R}. \label{P2Ieq_rate constraint}
		\end{align}
Our remaining task is to optimize $\bm{V}_{\mathrm{RF}}$ with $\{\bm{Q}_k\}_{k=1}^{K}$ and $\{\bm{U}_k\}_{k=1}^{K}$ being fixed.
Let $\bm{D}_{1,k}=\bm{H}_k^H\bm{Q}_k\bm{U}_k\bm{Q}_k^H\bm{H}_k$, $\bm{D}_{2,k}=\bm{V}_{\mathrm{BB},k}\bm{U}_k\bm{Q}_k^H\bm{H}_k$, $\sum_{k=1}^K(\bm{R}_{\mathrm{BB},k}^T\otimes\tilde{\bm{A}})=\bm{\Upsilon}_1$, $\sum_{k=1}^K(\bm{R}_{\mathrm{BB},k}^T\otimes\bm{I}_{N_\mathrm{T}})=\bm{\Upsilon}_2$, $\frac{1}{K}\sum_{k=1}^K(\bm{R}_{\mathrm{BB},k}^T\otimes\bm{D}_{1,k})=\bm{\Upsilon}_3$, and $\bar{\eta}=\frac{1}{K}(\sum_{k=1}^K(\log|\bm{U}_k|-\mathrm{tr}(\bm{U}_k)-\bar{\sigma}_{\mathrm{C}}^2\mathrm{tr}(\bm{U}_k\bm{Q}_k^H\bm{Q}_k)+N_{\mathrm{S},k}))$, where $\bm{\Upsilon}_1$, $\bm{\Upsilon}_2$, and $\bm{\Upsilon}_3$ are all positive semi-definite matrices. The problem of optimizing $\bm{V}_{\mathrm{RF}}$ for (P2-I-eqv) is thereby equivalently expressed as
		\begin{align}
			\!\!\!\mbox{(P2-I-eqv')}\, \underset{\bm{v}}{\max}\,\,& 
			\bm{v}^H\bm{\Upsilon}_1\bm{v}\label{P2Ieqvm_obj}\\
			\mbox{s.t.}\quad &\bm{v}^H\bm{\Upsilon}_2\bm{v}\leq P \label{P2Ieqvm_power constraint} \\   
			&-\bm{v}^H\bm{\Upsilon}_3\bm{v}+2\mathfrak{Re}(\Tilde{\bm{c}}^T\bm{v})\geq \Bar{R}-\bar{\eta} \label{P2Ieqvm_rate constraint}\\
			& \bm{v}^H \bm{\Lambda}_m\bm{v}= 1, \, m=1,...,N_{\mathrm{T}}N_{\mathrm{RF},\mathrm{T}},
		\end{align}
where $\bm{c}_k=\mathrm{vec}(\bm{D}_{2,k}^T)$, $\Tilde{\bm{c}}=\frac{1}{K}\sum_{k=1}^K\bm{c}_k$. Although (P2-I-eqv') is still non-convex, we apply the FPP-SCA technique by iteratively solving the following problem with a given local point $\bar{\bm{v}}$:
		\begin{align}
			\!\!\!\!\!\mbox{(P2-I-F)}\!\!\!\!\!\!\!\! \underset{\substack{\bm{v}:(\ref{P2Ieqvm_power constraint}),(\ref{P2Ieqvm_rate constraint}), t,\\ r\geq 0,\bm{p}\succeq\bm{0}, \bm{w}\succeq\bm{0}}}{\max}\!\!\!\!\!\!& 
			t-\epsilon r-\epsilon \|\bm{p}\|_1-\epsilon\|\bm{w}\|_1 \label{P2If_obj}\\
			\mbox{s.t.}\quad\ & 2\mathfrak{Re}\{\bar{\bm{v}}^H\bm{\Upsilon}_1\bm{v} \}\!-\!\bar{\bm{v}}^H\bm{\Upsilon}_1\bar{\bm{v}}
			\geq t-r\\
			& \bm{v}^H \bm{\Lambda}_m\bm{v}\leq 1+p_m,\quad \forall m\\
			& 2\mathfrak{Re}\big\{\bar{\bm{v}}^H\bm{\Lambda}_m\bm{v}\big\}\!\!-\!\bar{\bm{v}}^H\bm{\Lambda}_m\bar{\bm{v}}\!\geq\! 1\!-\!w_m,  \forall m.
		\end{align}
Note that (P2-I-F) is a convex optimization problem which can be efficiently solved via the interior-point method \cite{CVX} or existing software such as CVX \cite{cvxtool}. Thus, with $\bar{\bm{v}}$ iteratively updated as the optimal $\bm{v}$ obtained by solving (P2-I-F) in the previous iteration, the FPP-SCA method is guaranteed to converge, and will converge to at least a KKT point to (P2-I-eqv') as long as the slack variables are converged to zero \cite{mehanna2014feasible}. 
\subsubsection{Optimal solution to $\{\bm{R}_{\mathrm{BB},k}\}_{k=1}^K$}
With given $\bm{V}_{\mathrm{RF}}$ and $\bm{W}_{\mathrm{RF}}^H$, the transmit digital beamforming in $\{\bm{R}_{\mathrm{BB},k}\}_{k=1}^K$ can be optimized via solving the following problem:
		\begin{align}
			\!\!\!\!\mbox{(P2-II)}\!\!\!\!\!\!\!\underset{\{\bm{R}_{\mathrm{BB},k}\succeq\bm{0}\}_{k=1}^K:(\ref{P2_rate constraint}),(\ref{P2_power constraint})}{\max}\! & \sum_{k=1}^K\mathrm{tr}\big(\tilde{\bm{A}}\bm{V}_{\mathrm{RF}}\bm{R}_{\mathrm{BB},k}\bm{V}_{\mathrm{RF}}^H\big). \label{P2II_obj}
		\end{align}
 (P2-II) is a convex problem, for which the optimal solution can be obtained via the interior-point method \cite{CVX}
or existing software such as CVX \cite{cvxtool}. \looseness=-1

\subsubsection{Optimal solution to each element in $\bm{W}_{\mathrm{RF}}^H$}
With given $\{\bm{R}_{\mathrm{BB},k}\}_{k=1}^K$ and $\bm{V}_{\mathrm{RF}}$, the subproblem for optimizing $\bm{W}_{\mathrm{RF}}^H$ has the same structure as the sensing-only case. Thus, the optimal solution to each element of $\bm{W}_{\mathrm{RF}}^H$ can be obtained via (\ref{wRF_solution}), where $\tilde{\bm{B}}$ is replaced by $\bar{\bm{B}}=\sum_{k=1}^K\int \Dot{\bm{M}}(\theta)\bm{V}_{\mathrm{RF}}\bm{R}_{\mathrm{BB},k}\bm{V}_{\mathrm{RF}}^H\Dot{\bm{M}}^H(\theta)p_{\Theta}(\theta)d\theta$.

\subsubsection{Overall AO-based algorithm for (P2)}
To summarize, the proposed AO-based algorithm for MIMO-OFDM ISAC iteratively optimizes $\bm{V}_{\mathrm{RF}}$ using the WMMSE transformation and the FPP-SCA technique, $\{\bm{R}_{\mathrm{BB},k}\}_{k=1}^K$ by solving the convex problem (P2-II), and $\bm{W}_{\mathrm{RF}}^H$ via the element-wise update (\ref{wRF_solution}). The algorithm is guaranteed to converge monotonically.
 The worst-case complexity of proposed algorithm can be shown to be $\mathcal{O}(N_{\mathrm{out}}(N_{\mathrm{T}}^2\bar{\omega}+N_{\mathrm{R}}N_{\mathrm
T}^2+N_{\mathrm{R}}N_{\mathrm{T}}N_{\mathrm{RF},\mathrm{R}}+K(N_{\mathrm{R}}^2\bar{\omega}+N_{\mathrm{R}}^2N_{\mathrm{T}}+N_{\mathrm{R}}N_{\mathrm{T}}N_{\mathrm{RF},\mathrm{T}}+N_{\mathrm{R}}N_{\mathrm{RF},\mathrm{T}}^2)+K^{3.5}N_{\mathrm{RF},\mathrm{T}}^7+N_{\mathrm{V}}(N_{\mathrm{in}}(N_{\mathrm{T}}N_{\mathrm{RF},\mathrm{T}})^{3.5}+K(N_{\mathrm{T}}^2N_{\mathrm{RF},\mathrm{T}}^2+N_{\mathrm{U}}^3+N_{\mathrm{U}}N_{\mathrm{T}}^2+N_{\mathrm{U}}^2N_{\mathrm{T}}+N_{\mathrm{U}}N_{\mathrm{T}}N_{\mathrm{RF},\mathrm{T}}+N_{\mathrm{U}}N_{\mathrm{RF},\mathrm{T}}^2))))$ where $N_{\mathrm{in}}$, $N_{\mathrm{V}}$, and $N_{\mathrm{out}}$ denote the numbers of FPP-SCA iterations, WMMSE iterations, and outer iterations, respectively.

\vspace{-4mm}
\section{Hybrid Beamforming Design under Fully-Connected Receiver Architecture} \label{sec:extension}
\vspace{-1mm}

	The previous sections have focused on the partially-connected receiver architecture shown in Fig. \ref{system model},  where each antenna is only connected to one RF chain. In this case, the inherent receive analog beamforming orthogonality brought by the constraints in (\ref{WRF_constraint}), i.e., $\bm{W}_{\mathrm{RF}}^H\bm{W}_{\mathrm{RF}}=\frac{N_{\mathrm{R}}}{N_{\mathrm{RF},\mathrm{R}}}\bm{I}_{N_{\mathrm{RF},\mathrm{R}}}$, enables tractable PCRB expressions and effective beamforming design. 
This section extends our study to the \emph{fully-connected architecture} \cite{sohrabi2016hybrid} where each antenna is connected to \emph{all} RF chains, and the analog combining matrix $\bm{W}_{\mathrm{RF}}^H$ is subjected to an element-wise unit-modulus constraint given by
\begin{equation}\label{WRF_fully-connected}
    |[\bm{W}_{\mathrm{RF}}^H]_{i,j}|=1,\quad \forall i,j.
\end{equation}
The PCRB under this architecture is extremely difficult to characterize. Specifically, the PCRB involves the integration of $\bm{C}_y^{-1}$, where $\bm{C}_y=\sigma_{\mathrm{S}}^2(\bm{I}_L\otimes\bm{W}_{\mathrm{RF}}^H)(\bm{I}_L\otimes\bm{W}_{\mathrm{RF}}^H)^H$ is the effective receiver noise covariance matrix. Under the new constraints in (\ref{WRF_fully-connected}), $\bm{W}_{\mathrm{RF}}^H\bm{W}_{\mathrm{RF}}$ is generally a dense positive semi-definite matrix  whose inversion is mathematically intractable. To overcome this new challenge, we propose a structured design of $\bm{W}_{\mathrm{RF}}^H$ which satisfies $\bm{W}_{\mathrm{RF}}^H\bm{W}_{\mathrm{RF}}=N_{\mathrm{R}}\bm{I}_{N_{\mathrm{RF},\mathrm{R}}}$ under (\ref{WRF_fully-connected}). With this additional orthogonality constraint, the effective receiver noise remains spatially white after analog combining. Consequently, the PCRB in estimating $\theta$ for the narrowband case is given by \looseness=-1
    \begin{align}\label{PCRB_FC}
        \!\!\!
        \mathrm{PCRB}_{\theta}^{\mathrm{FC}}\!=\!\frac{1}{\mathbb{E}_\theta\Big[\Big(\frac{\partial\ln(p_{\Theta}(\theta))}{\partial\theta}\Big)^2\Big]\!\!+\!\!\frac{2L\gamma}{N_{\mathrm{R}}\sigma_{\mathrm{S}}^2}\mathrm{tr}(\Tilde{\bm{B}}\bm{W}_{\mathrm{RF}}\bm{W}_{\mathrm{RF}}^H)
        }\!,
    \end{align}
where $\tilde{\bm{B}}=\int\Dot{\bm{M}}(\theta)\bm{V}_{\mathrm{RF}}\bm{R}_{\mathrm{BB}}\bm{V}_{\mathrm{RF}}^H\Dot{\bm{M}}^H(\theta)p_{\Theta}(\theta)d\theta$. To solve the PCRB minimization problem, we propose an effective DFT codebook-based approach. Specifically, we select the rows of $\bm{W}_{\mathrm{RF}}^H$ from a normalized DFT matrix $\bm{F}=[\bm{f}_1,...,\bm{f}_{N_{\mathrm{R}}}] \in \mathbb{C}^{N_{\mathrm{R}}\times N_{\mathrm{R}}}$, where $[\bm{f}_i]_j=e^{\frac{-j2\pi(i-1)(j-1)}{N_{\mathrm{R}}}}$. Then, the optimization problem under the fully-connected receiver architecture is formulated as
		\begin{align}
			\mbox{(P3)}
            \!\!\!\!\!\!\!\underset{\substack{\bm{V}_{\mathrm{RF}},\bm{R}_{\mathrm{BB}}\succeq\bm{0},\\ \bm{W}_{\mathrm{RF}}^H:(\ref{WRF_fully-connected}),\bm{q},\\(\ref{P1_rate constraint})-(\ref{P1_unit-modulus constraint1})}}{\max} & \sum_{n=1}^{N_{\mathrm{RF},\mathrm{R}}}\bm{f}_{q_n}^H\tilde{\bm{B}}\bm{f}_{q_n} \label{P3_obj}\\
            \mbox{s.t.}\quad\quad\ &\!\! q_n\!\in\!\{1,...,N_{\mathrm{R}}\}, \forall n,\  q_n\!\neq\! q_m,  \forall m\!\neq\! n,
		\end{align}
where $\bm{q}\in\mathbb{R}^{N_{\mathrm{RF},\mathrm{R}}\times 1}$ represents the indices of the selected rows in the DFT matrix. 
Note that (P3) can be similarly dealt with using the AO-based algorithm by iteratively optimizing $\bm{V}_{\mathrm{RF}}$ in a similar manner as that in Section \ref{section_narrowband_VRF}, $\bm{R}_{\mathrm{BB}}$ via (\ref{RBB_solution}), and the selected indices in $\bm{q}$ whose optimal solution can be shown to be $\bm{q}^{\star}=[\kappa(1),...,\kappa(N_{\mathrm{RF},\mathrm{R}})]^T$ with $\kappa(\cdot)$ denoting the permutation of $\{1,...,N_{\mathrm{R}}\}$ that sorts the diagonal elements of $\bm{F}^H\tilde{\bm{B}}\bm{F}$ in a descending order, i.e., $[\bm{F}^H\tilde{\bm{B}}\bm{F}]_{\kappa(1),\kappa(1)}\geq,...,\geq[\bm{F}^H\tilde{\bm{B}}\bm{F}]_{\kappa(N_{\mathrm{R}}),\kappa(N_{\mathrm{R}})}$.
Monotonic convergence is guaranteed for the AO-based algorithm. The above results can also be readily generalized to the MIMO-OFDM case, which is omitted for brevity.
\looseness=-1

\vspace{-5mm}
\section{Numerical Results}\label{sec_numerical results}
\vspace{-2mm}

\begin{figure}[t]
		\centering
		\vspace{-4mm}
		\includegraphics[width=0.46\textwidth]{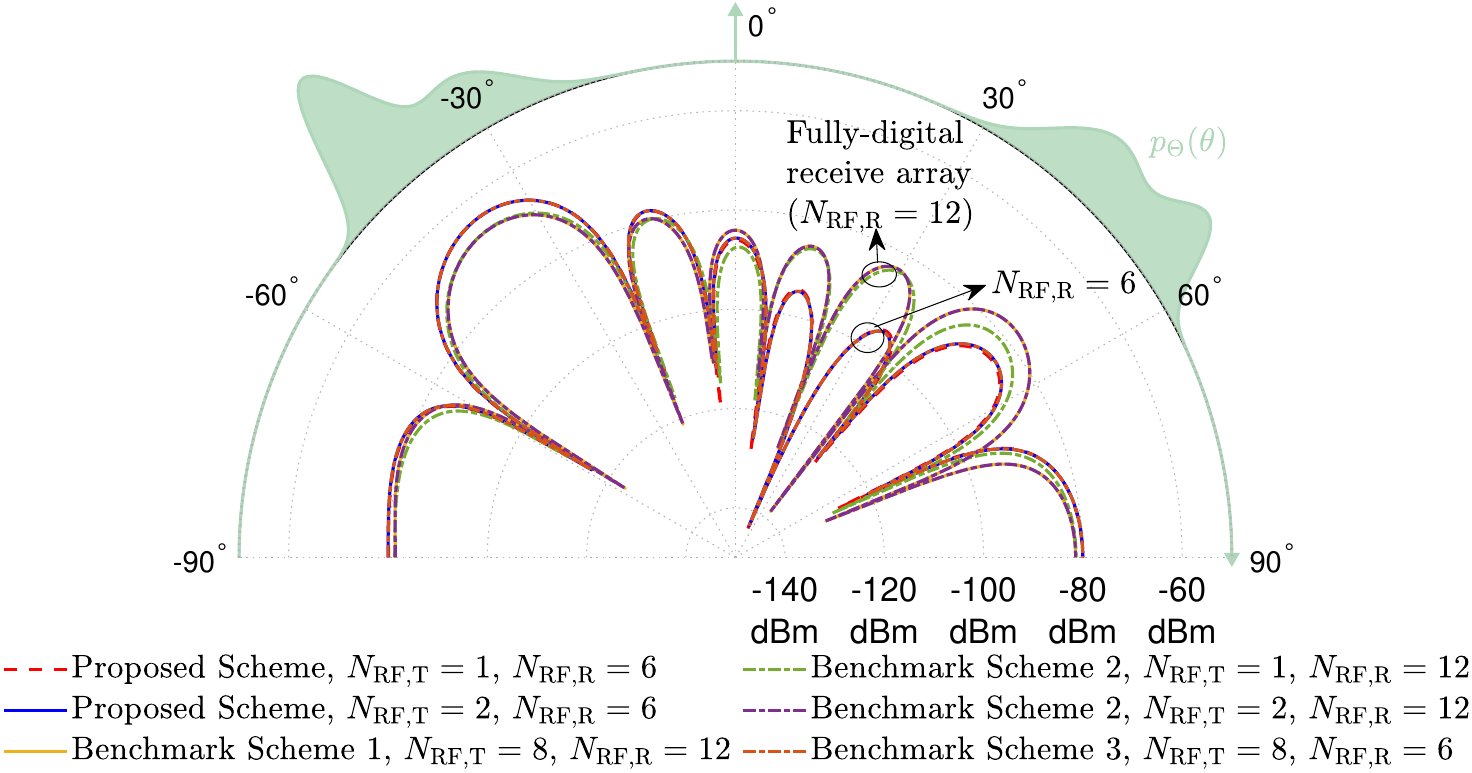}
		\vspace{-4mm}
		\caption{BS received power pattern over $\theta$ for MIMO sensing.}
		\label{receive_power_pattern_sensing}
	\end{figure}

In this section, we provide numerical results to evaluate the performance of our proposed hybrid beamforming designs. Unless specified otherwise, we set $N_{\mathrm{T}}=8$, $N_{\mathrm{R}}=12$, $N_{\mathrm{U}}=6$, $P=30$ dBm, $\sigma_{\mathrm{C}}^2=\sigma_{\mathrm{S}}^2=-90$ dBm, $\bar{\sigma}_{\mathrm{C}}^2=\bar{\sigma}_{\mathrm{S}}^2=-90$ dBm, $N_{\mathrm{RF},\mathrm{T}}=3$, $N_{\mathrm{RF},\mathrm{R}}=6$, $\alpha\sim\mathcal{CN}(0,2\times 10^{-12})$, and a rate target of $\hat{R}=4.5$ bits/second/Hz (bps/Hz) or equivalently $\bar{R}=\log(2)\hat{R}=3.119$ nps/Hz.
We consider a ULA layout for all antenna arrays with half-wavelength spacing, which yields $\bm{a}(\theta)=[e^{\frac{-j\pi (N_{\mathrm{T}}-1)\sin\theta}{2}},...,e^{\frac{j\pi (N_{\mathrm{T}}-1)\sin\theta}{2}}]^T$ and $\bm{b}(\theta)=[e^{\frac{-j\pi (N_{\mathrm{R}}-1)\sin\theta}{2}},...,e^{\frac{j\pi (N_{\mathrm{R}}-1)\sin\theta}{2}}]^T$.
In practice, the angle distribution of a target is usually concentrated around one or multiple nominal angles. Therefore, we assume that the PDF of $\theta$ follows a Gaussian mixture model, which is the
weighted summation of $I\geq 1$ Gaussian PDFs and is given by $
		p_{\Theta}(\theta)=\sum_{i=1}^I\frac{p_i}{\sqrt{2\pi}\sigma_i}e^{-\frac{(\theta-\theta_i)^2}{2\sigma_i^2}}$. Specifically, $\theta_i\in[-\frac{\pi}{2}, \frac{\pi}{2})$ and $\sigma_i^2$ are the mean and variance of the $i$-th Gaussian PDF, respectively, and $p_i$ denotes the weight that satisfies $\sum_{i=1}^Ip_i=1$. We further set $I=4$; $\theta_1=-0.74$, $\theta_2=-0.54$, $\theta_3=-0.75$, $\theta_4=0.95$; $\sigma_1^2=10^{-2.5}$, $\sigma_2^2=10^{-2}$, $\sigma_3^2=10^{-2}$, $\sigma_4^2=10^{-2.5}$; and $p_1=0.31$, $p_2=0.24$, $p_3=0.28$, $p_4=0.17$. The PDF of $\theta$ is illustrated in Fig. \ref{receive_power_pattern_sensing}. \looseness=-1
   	
	We assume the communication user is located at a distance of $r_{\mathrm{U}}=400$ m and an angle of $\theta_{\mathrm{U}}=0.36$ with respect to the BS. For narrowband communication, we consider a Rician  
	fading  channel model with $\bm{H}=\sqrt{\beta_{\mathrm{C}}/(K_{\mathrm{C}}+1)}(\sqrt{K_{\mathrm{C}}}\bm{H}_{\mathrm{LoS}}+\bm{H}_{\mathrm{NLoS}})$, where $\beta_{\mathrm{C}}=\frac{\beta_0}{r_{\mathrm{U}}^{3.5}}$ denotes the path power gain with $\beta_0=-30$ dB and $K_{\mathrm{C}}=-8$ dB denotes the Rician factor. The LoS component is set as $\bm{H}_{\mathrm{LoS}}=\bm{b}_{\mathrm{U}}(\theta_{\mathrm{U}})\bm{a}(\theta_{\mathrm{U}})^H$, where $\bm{b}_{\mathrm{U}}(\theta_{\mathrm{U}})$ denotes the steering vector of the user receive antenna array. The non-LoS (NLoS) component follows a distribution of $[\bm{H}_{\mathrm{NLoS}}]_{m,n}\!\sim\! \mathcal{CN}(0,1)$, $\forall m, n$. 
	For wideband communication, let $L_{\mathrm{D}}=8$ denote the number of delayed taps in the time-domain impulse responses. $\Bar{\bm{H}}_l\in \mathbb{C}^{N_{\mathrm{U}}\times N_{\mathrm{T}}}$, $l\in\{0,...,L_{\mathrm{D}}-1\}$ denotes the time-domain channel matrix at each $l$-th tap, where $[\bar{\bm{H}}_l]_{i,j}\sim\mathcal{CN}(0,\frac{\bar{\beta}_{\mathrm{C}}}{L_{\mathrm{D}}})$ with $\bar{\beta}_{\mathrm{C}}=\frac{\beta_0}{r_{\mathrm{U}}^{2.8}}$ denoting the path power gain. The frequency-domain channel matrix at each $k$-th sub-carrier is given by $\bm{H}_k=\sum_{l=0}^{L_{\mathrm{D}}-1}\Bar{\bm{H}}_le^{-j2\pi(k-1)l/K}$, $k=1,..., K$. The number of sub-carriers is set as $K=16$.\looseness=-1

    \begin{figure}[t]
	   \centering
	   \includegraphics[width=0.46\textwidth]{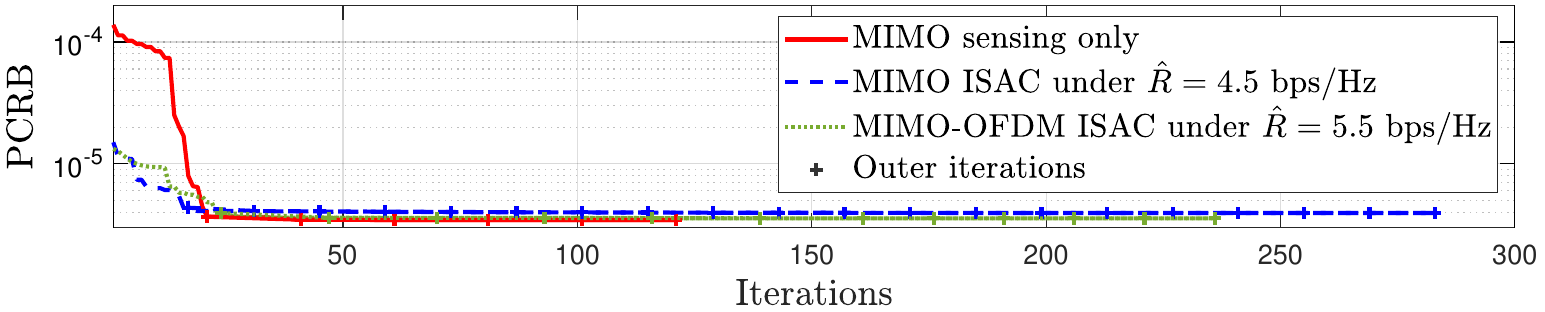}
	   \vspace{-4mm}
	   \caption{Convergence behavior of the proposed AO-based algorithms.}
	   \label{Convergence behavior}
    \end{figure}
	First, we show the convergence behavior of our proposed AO-based algorithms for different systems in Fig. \ref{Convergence behavior}, where we  
	set $N_{\mathrm{RF},\mathrm{T}}=1$ for MIMO sensing, and $N_{\mathrm{RF},\mathrm{T}}=2$ for MIMO ISAC and MIMO-OFDM ISAC with $\hat{R}=4.5$ bps/Hz and $\hat{R}=5.5$ bps/Hz, respectively. It is observed  that all the 	
    proposed AO-based algorithms converge monotonically and fastly within a few outer iterations. Then, we evaluate the performance of the proposed algorithms in the following.\looseness=-1 

    \begin{figure}[t]
		\centering
		\includegraphics[width=0.46\textwidth]{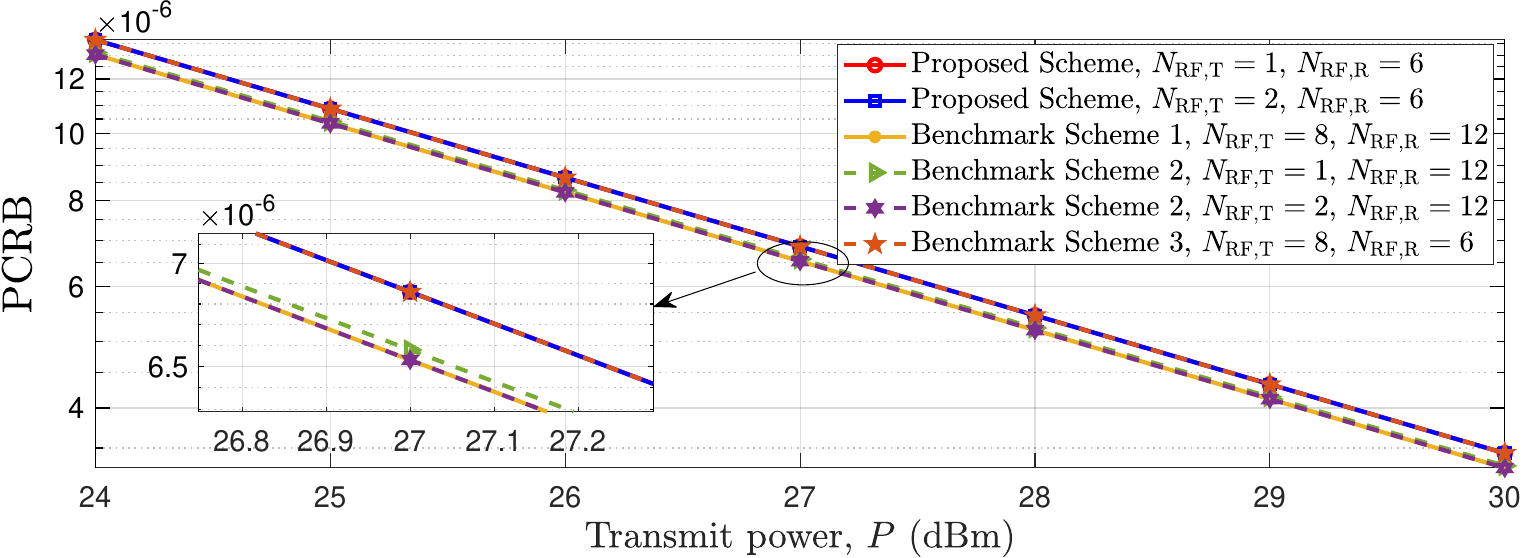}
		\vspace{-4mm}
		\caption{MIMO sensing performance under different array structures.}
		\label{P_VS_PCRB_BothHB_Sensing}
	\end{figure}
	\begin{figure}[H]
    \vspace{-1mm}
		\centering
		\includegraphics[width=0.44\textwidth]{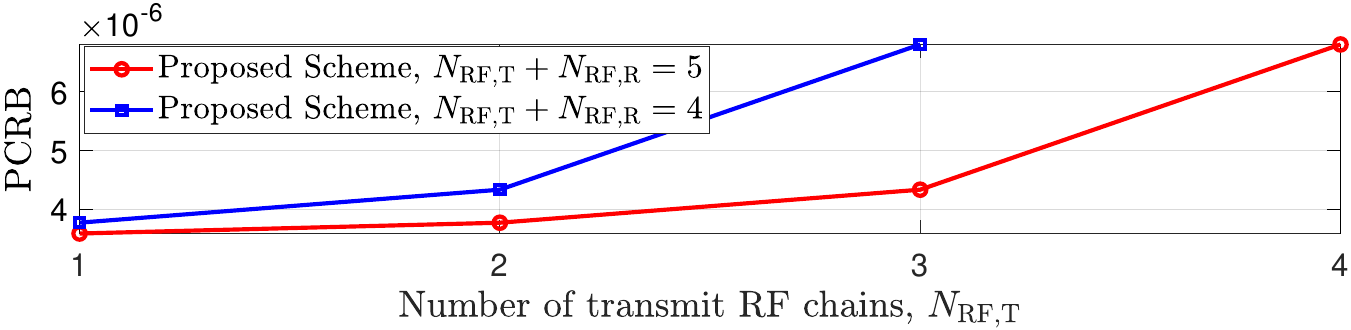}
		\vspace{-4mm}
		\caption{PCRB versus the number of transmit RF chains for MIMO sensing.}
		\label{TxRF_vs_PCRB_sensing}
	\end{figure}

\vspace{-9mm}
\subsection{Performance with Different Numbers of RF Chains}
\vspace{-1mm}  
	In this subsection, we evaluate the performance of the proposed hybrid beamforming algorithms by comparing them with the optimal fully-digital beamforming designs, and examine the effect of the number of transmit/receive RF chains on the performance. We consider the following benchmark schemes:
	\begin{itemize}
		\item {\bf{Benchmark Scheme 1: Optimal fully-digital beamforming}.} In this scheme, both the transmitter and receiver adopt fully-digital arrays. The optimal fully-digital beamforming is obtained by solving (P0-FD), (P1), or (P2) with $\bm{V}_{\mathrm{RF}}=\bm{I}_{N_{\mathrm{T}}}$ and $\bm{W}_{\mathrm{RF}}^H=\bm{I}_{N_\mathrm{R}}$, which are convex problems.\looseness=-1
		\item {\bf{Benchmark Scheme 2: Fully-digital receive array}.} In this scheme, the receiver adopts a fully-digital array with $\bm{W}_{\mathrm{RF}}^H=\bm{I}_{N_\mathrm{R}}$. (Each element in) $\bm{V}_{\mathrm{RF}}$ and $\bm{R}_{\mathrm{BB}}$ are iteratively optimized via our proposed AO-based framework.
		\item {\bf{Benchmark Scheme 3: Fully-digital transmit array}.} In this scheme, the transmitter adopts a fully-digital array with $\bm{V}_{\mathrm{RF}}=\bm{I}_{N_{\mathrm{T}}}$. Each element in $\bm{W}_{\mathrm{RF}}^H$ and $\bm{R}_{\mathrm{BB}}$ are iteratively optimized via our proposed AO-based framework. \looseness=-1
	\end{itemize}

\subsubsection{MIMO Sensing System}
	Firstly, we consider the MIMO sensing system. Fig. \ref{P_VS_PCRB_BothHB_Sensing} shows the PCRB versus transmit power for the proposed scheme and benchmark schemes under various numbers of RF chains. Fig. \ref{receive_power_pattern_sensing} shows the average received power pattern at the BS receiver versus the target's angle $\theta$, which is given by
	$\mathcal{P}_{\mathrm{R}}(\theta)= \mathbb{E}[|\alpha|^2]\|\bm{W}_{\mathrm{RF}}^H\bm{b}(\theta)\bm{a}(\theta)^H\bm{V}_{\mathrm{RF}}\bm{V}_{\mathrm{BB}}\|_{\mathrm{F}}^2$,
	together with the PDF of $\theta$, $p_{\Theta}(\theta)$. We start by comparing the schemes with the same number of receive RF chains, $N_{\mathrm{RF},\mathrm{R}}$. It is observed that Benchmark Scheme 2 with fully-digital receive array achieves exactly the same performance and BS received power pattern as the optimal fully-digital beamforming in Benchmark Scheme 1 when $N_{\mathrm{RF},\mathrm{T}}=2$, and very similar performance and power pattern when $N_{\mathrm{RF},\mathrm{T}}=1$. This thus validates our analytical results in Proposition \ref{prop_solution1} and indicates that \emph{$1$ transmit} \emph{RF chain} suffices to achieve near-optimal MIMO sensing with a fully-digital BS receiver, despite the continuous range of the target's possible angle. Similarly, under $N_{\mathrm{RF},\mathrm{R}}=6$, Benchmark Scheme 3 performs almost the same as the proposed scheme with $N_{\mathrm{RF},\mathrm{T}}=2$ or $N_{\mathrm{RF},\mathrm{T}}=1$. This further validates the effectiveness of our proposed AO-based framework. \looseness=-1
	
	\begin{figure}[t]
		\centering
		\includegraphics[width=0.46\textwidth]{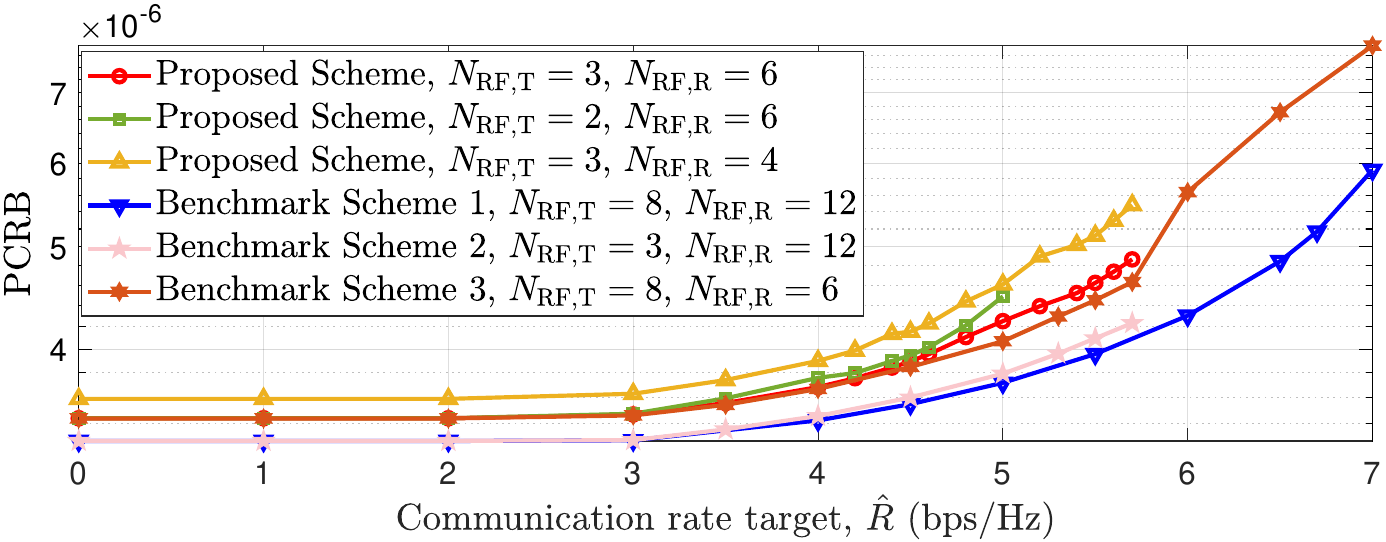}
		\vspace{-4mm}
		\caption{MIMO ISAC performance under different array structures.}
		\label{R_VS_PCRB_ISAC_narrow_differentRF}
	\end{figure} 
	\begin{figure}[t]
		\centering
		\vspace{-5mm}
		\includegraphics[width=0.45\textwidth]{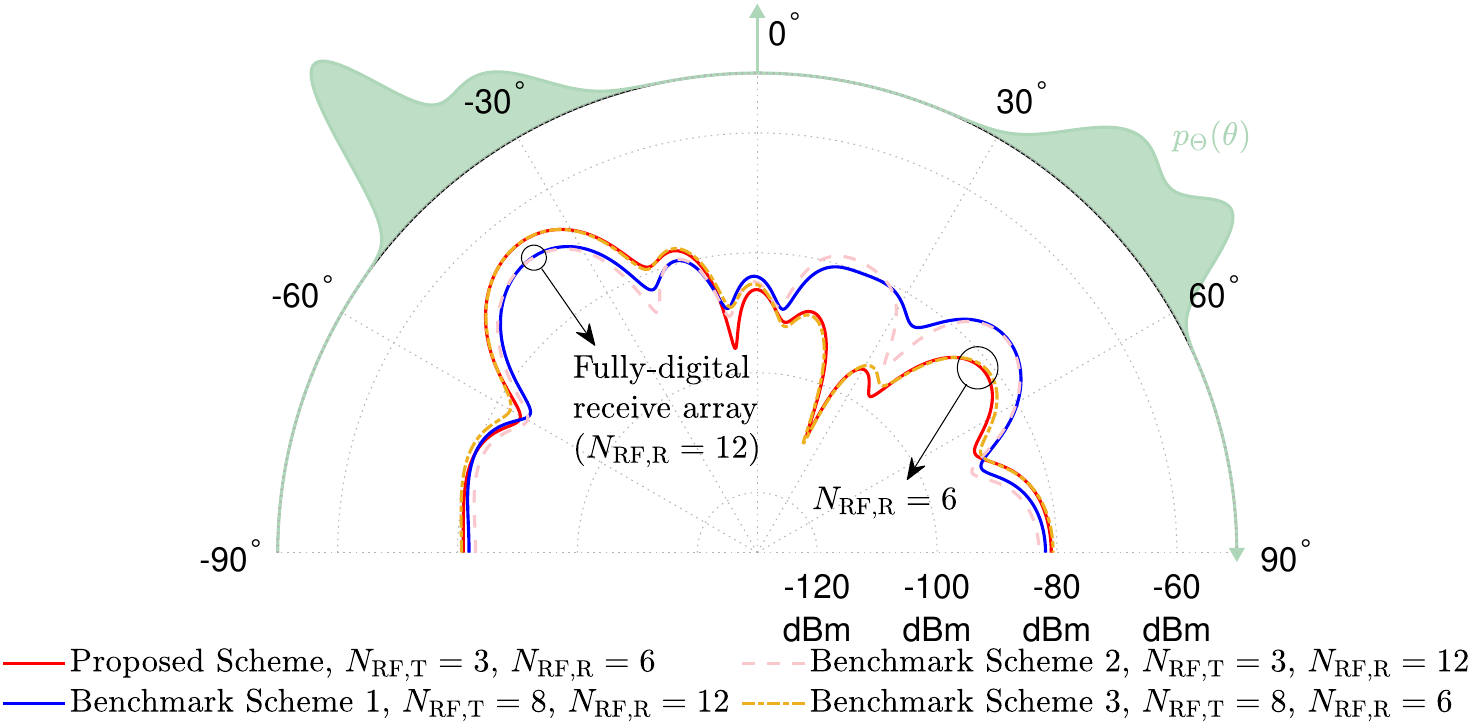}
		\vspace{-4mm}
		\caption{BS received power pattern over $\theta$ for MIMO ISAC.}
		\label{receive_power_pattern_ISAC}
	\end{figure}
	
	Then, we compare the schemes with the same number of transmit RF chains, $N_{\mathrm{RF},\mathrm{T}}$. It is observed from Fig. \ref{P_VS_PCRB_BothHB_Sensing} that the PCRB performance degrades as $N_{\mathrm{RF},\mathrm{R}}$ decreases. Moreover, it is observed from Fig. \ref{receive_power_pattern_sensing} that there are two ranges of $\theta$ that accumulate most of the probability density. With fewer receive RF chains, the BS received power is more focused over one of these ranges, instead of being spread more evenly between the two ranges, which explains the performance loss.\looseness=-1

	\begin{figure}[t]
		\centering
		\includegraphics[width=0.46\textwidth]{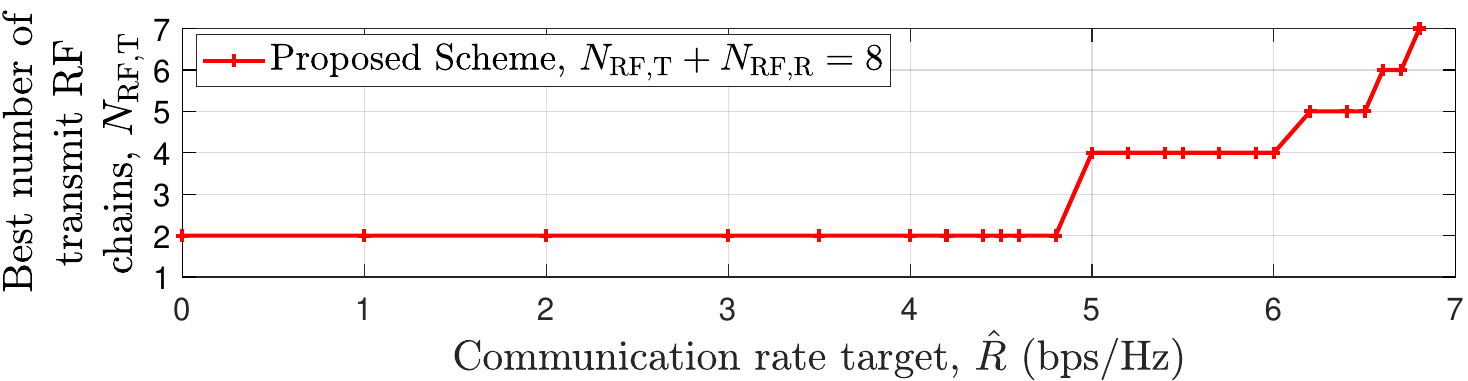}
		\vspace{-4mm}
		\caption{Best number of transmit RF chains for MIMO ISAC.}
		\label{OptimalTxRF_vs_R_ISAC}
	\end{figure} 
	\begin{figure}[t]
		\centering
		\vspace{-4mm}
		\includegraphics[width=0.46\textwidth]{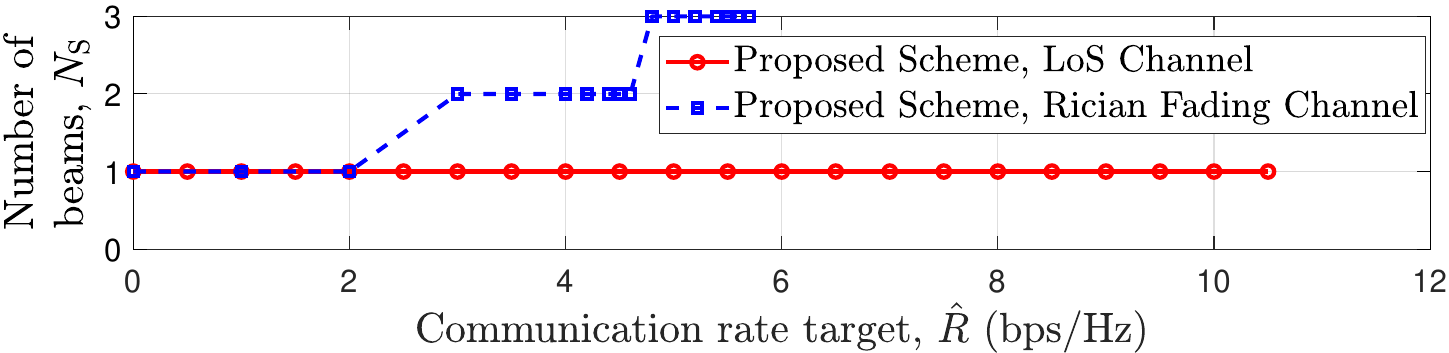}
		\vspace{-4mm}
		\caption{Number of dual-functional beams, $N_{\mathrm{S}}$, for MIMO ISAC.}
		\label{R_VS_NS_ISAC_Both_HB}
	\end{figure}
	
	Lastly, it is worth noting that the above results already imply that the limitation of RF chains at the receiver may have a higher effect on the MIMO sensing performance compared to that at the transmitter. Moreover, it is observed from Fig. \ref{P_VS_PCRB_BothHB_Sensing} and Fig. \ref{receive_power_pattern_sensing} that Benchmark Scheme 3 with fully-digital transmit array and $6$ receive RF chains is significantly outperformed by Benchmark Scheme 2 with fully-digital receive array and $2$ transmit RF chains. To further examine this effect, we consider a given budget on the total number of RF chains at the transmitter and the receiver, and show in Fig. \ref{TxRF_vs_PCRB_sensing} the PCRB versus the number of transmit RF chains, $N_{\mathrm{RF},\mathrm{T}}$. It is observed that the PCRB increases as $N_{\mathrm{RF},\mathrm{T}}$ increases, and the optimal allocation strategy is to only put \emph{a single RF chain at the transmitter}, and all the remaining ones at the receiver. This is consistent with our observations from Figs. \ref{P_VS_PCRB_BothHB_Sensing} and \ref{receive_power_pattern_sensing}. \looseness=-1

\subsubsection{MIMO ISAC System}
	Next, we consider the MIMO ISAC system. Fig. \ref{R_VS_PCRB_ISAC_narrow_differentRF} shows the PCRB versus rate target $\hat{R}$ for the proposed scheme and benchmark schemes under various numbers of RF chains. It is observed that both the PCRB performance and the maximum supportable rate improve as the number of transmit RF chains increases, and the PCRB performance also improves as the number of receive RF chains increases. The proposed AO-based algorithm performs closely to the optimal fully-digital design in Benchmark Scheme 1, validating its effectiveness. Moreover, Fig. \ref{receive_power_pattern_ISAC} shows the  power pattern $\mathcal{P}_{\mathrm{R}}(\theta)$ at the BS receiver under $\hat{R}=4.5$ bps/Hz, where all schemes are observed to focus high power over highly-probable angles. However, having  fewer receive RF chains results in less balanced power distribution across possible angles, indicating reduced beamforming accuracy. \looseness=-1
	
	Notice from Fig. \ref{receive_power_pattern_ISAC} that reducing the number of RF chains at the receiver leads to more significant change in the power pattern. To further examine this effect, we consider a fixed budget of the total number of RF chains at the BS transceivers given by $N_{\mathrm{RF},\mathrm{T}}+N_{\mathrm{RF},\mathrm{R}}=8$, and show the \emph{best number of transmit RF chains, $N_{\mathrm{RF},\mathrm{T}}$}, versus the communication rate target $\hat{R}$ in Fig. \ref{OptimalTxRF_vs_R_ISAC}. Since the receiver employs a partially-connected hybrid array, the feasible set of $N_{\mathrm{RF},\mathrm{R}}$ is $\{1,2,3,4,6,12\}$, which yields the feasible set of $N_{\mathrm{RF},\mathrm{T}}$ as $\{2,4,5,6,7\}$. It is observed that in the low-to moderate rate regimes, it is optimal to allocate the \emph{smallest number of RF chains to the transmitter}. Although this result has also been observed for MIMO sensing, it is particularly interesting for MIMO ISAC as the transmit RF chains affect both communication and sensing. The reason behind may lie in the strong impact of the number of receive RF chains on the sensing performance, which still dominates in ISAC under low-to-moderate rate requirement. On the other hand, as the rate target increases, more data streams need to 
    be spatial multiplexed in the MIMO system to meet the high rate target, which requires the number of transmit RF chains to be increased. This further unveils the non-trivial trade-off 
    between sensing and communication in ISAC.
	
	Finally, Fig. \ref{R_VS_NS_ISAC_Both_HB} evaluates the number of dual-functional beams, $N_{\mathrm{S}}$ (i.e., rank of the transmit covariance matrix) versus the rate target. We consider $N_{\mathrm{RF},\mathrm{T}}=3$, $N_{\mathrm{RF},\mathrm{R}}=6$, and both the Rician fading communication channel and a new LoS communication channel, which yield the rank of the effective MIMO communication channel to be $3$ and $1$, respectively. It is observed that only one data stream is needed under the LoS channel, which shows that one dual-functional beam is sufficient under LoS channels for both the target and the communication user. On the other hand, as the rate target increases, the number of dual-functional beams increases under the Rician fading channel, and is upper bounded by the rank of the effective MIMO channel.

\subsubsection{MIMO-OFDM ISAC System} 

\begin{figure}[t]
		\centering
		\includegraphics[width=0.45\textwidth]{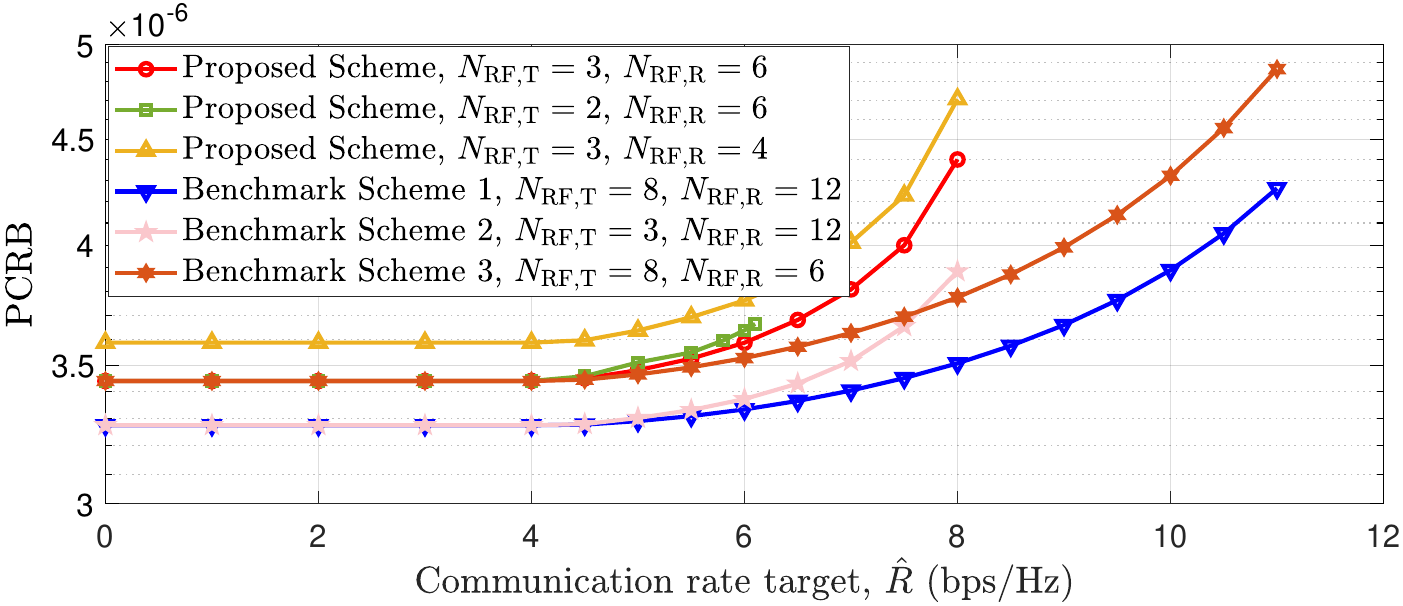}
		\vspace{-4mm}
		\caption{MIMO-OFDM ISAC performance under different array structures.}
		\label{R_VS_PCRB_OFDM_differentRF}
	\end{figure}

	In Fig. \ref{R_VS_PCRB_OFDM_differentRF}, we show the PCRB versus the communication rate target $\hat{R}$ for the proposed scheme and benchmark schemes with different numbers of RF chains. It is observed that as the number of transmit/receive RF chain decreases, the performance deteriorates more significantly compared to its narrowband counterpart, due to the higher\looseness=-1
    
	\begin{figure}[t]
		\centering
		\vspace{-4mm}
		\includegraphics[width=0.47\textwidth]{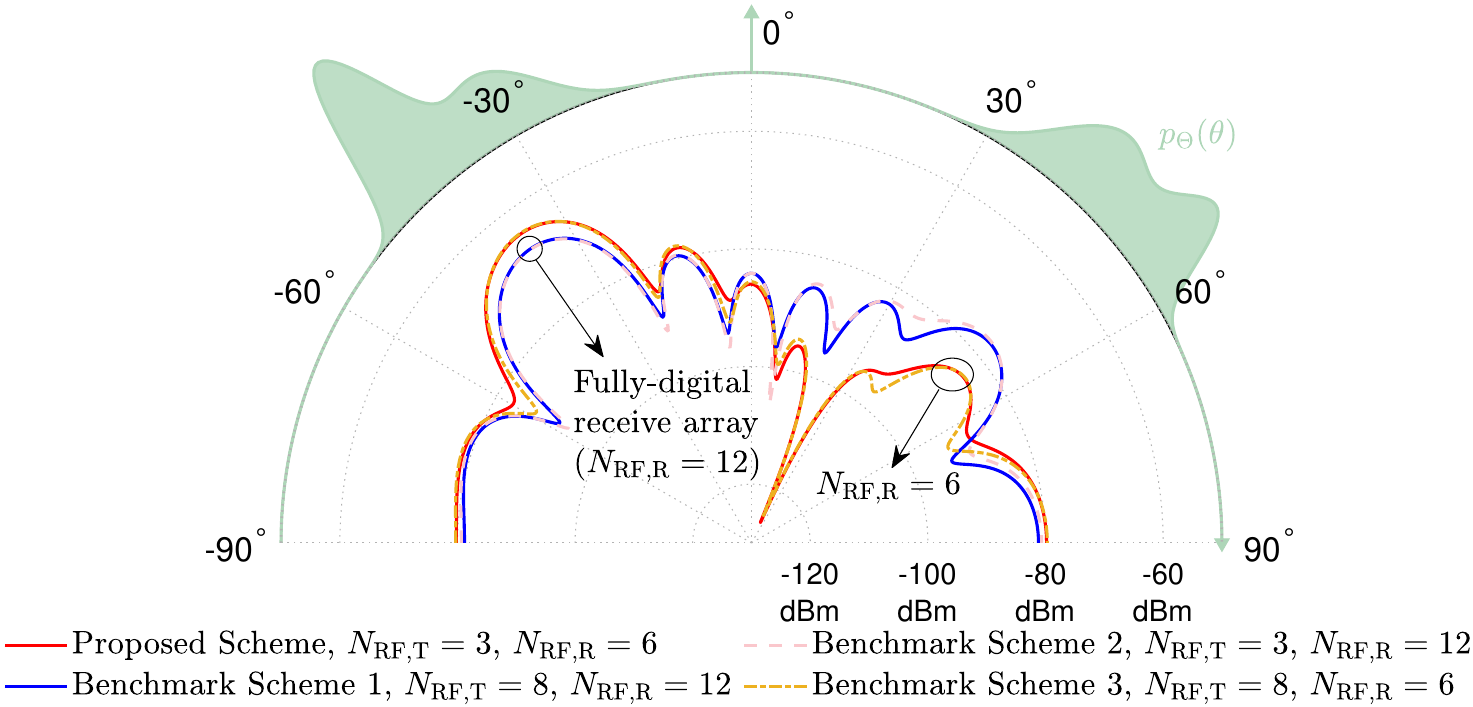}
		\vspace{-4mm}
		\caption{BS received power pattern over $\theta$ for MIMO-OFDM ISAC.}
		\label{receive_power_pattern_OFDM}
	\end{figure}
	\begin{figure}[H]
		\centering
		\includegraphics[width=0.46\textwidth]{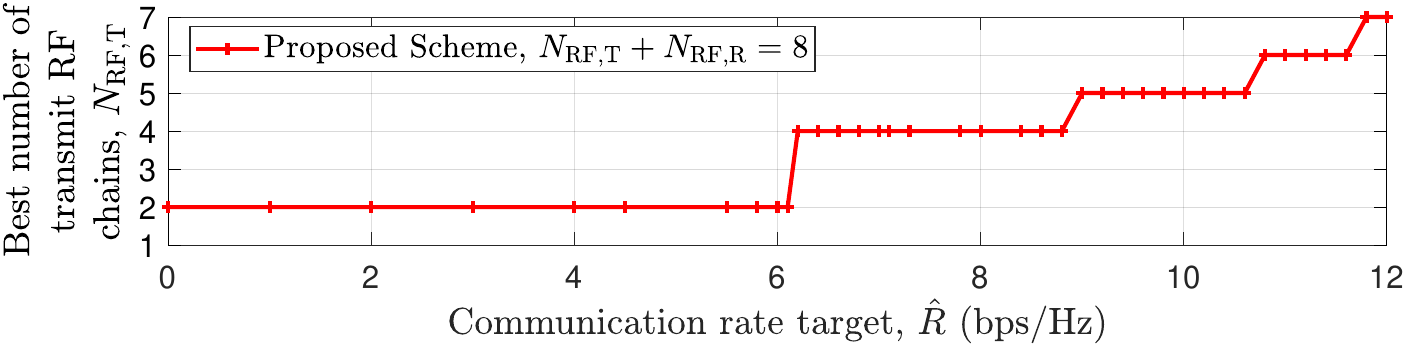}
		\vspace{-4mm}
		\caption{Best number of transmit RF chains for MIMO-OFDM ISAC.}
		\label{OptimalTxRF_vs_R_OFDM}
	\end{figure} 

    \vspace{-3mm}
    \noindent impact of limited RF chains on ISAC over multiple sub-carriers.  Fig. \ref{receive_power_pattern_OFDM} further shows the total power pattern over all sub-carriers at the BS receiver defined as $\mathcal{P}^{\mathrm{OFDM}}_{\mathrm{R}}(\theta)=\sum_{k=1}^K\mathbb{E}[|\alpha|^2]\|\bm{W}_{\mathrm{RF}}^H\bm{b}(\theta)\bm{a}(\theta)^H\bm{V}_{\mathrm{RF}}\bm{V}_{\mathrm{BB},k}\|_{\mathrm{F}}^2$ with $\hat{R}=5.5$ bps/Hz. It is also observed that the reduction of receive RF chains has a higher impact on the power pattern compared to its transmit counterpart.  \looseness=-1
	
	With a fixed total number of RF chains at the BS transmitter and BS receiver, Fig. \ref{OptimalTxRF_vs_R_OFDM} shows the best number of transmit RF chains, $N_{\mathrm{RF},\mathrm{T}}$, versus the communication rate target $\hat{R}$. It is also observed that the best number of transmit RF chains increases as the rate target increases.
	In the low-to-moderate rate regimes, the optimal allocation is to assign the \emph{smallest number of RF chains to the transmitter}, since the number of receive RF chains has a more obvious impact on the sensing performance.\looseness=-1

\vspace{-3mm}
\subsection{Performance Comparison with Other Designs}
\vspace{-1mm}

	In this subsection, we compare the performance of our proposed AO-based hybrid beamforming designs for MIMO ISAC and MIMO-OFDM ISAC against various other designs. Specifically, we consider the following benchmark schemes.	
	\begin{itemize}
		\item {\bf{Benchmark Scheme 4: Random phase based design}.} In this scheme, we randomly generate $N_{\mathrm{rand}}$ independent realizations of $\bm{V}_{\mathrm{RF}}$ and $\bm{W}_{\mathrm{RF}}^H$ under a uniform distribution for the phase of each unit-modulus element therein. Then, we obtain the optimal $\bm{R}_{\mathrm{BB}}$ for each realization, and select the best realization which yields the best objective value.\looseness=-1
		\item {\bf{Benchmark Scheme 5: Direction alignment based design}.} In this scheme, the transmit analog beamforming matrix is designed to align with the directions of the communication user and highly-probable locations of the target. Specifically, each  $j$-th column of $\bm{V}_{\mathrm{RF}}$ is designed as $\bm{a}(\theta_{\mathrm{U}})$ for $j=1$ and $\bm{v}_{\mathrm{RF},j}=\bm{a}(\theta_{j,\mathrm{max}})$ otherwise,
		where $\theta_{j,\mathrm{max}}$ is the mean value of the Gaussian PDF in the Gaussian mixture model with the $(j-1)$-th largest weight $p_i$. Note that this is similar to the heuristic design in \cite{liu2020joint1}. $\bm{W}_{\mathrm{RF}}^H$ and $\bm{R}_{\mathrm{BB}}$ are then designed under the proposed AO-based framework.
		\item {\bf{Benchmark Scheme 6: Partial prior information based design}.} In this scheme, we aim to minimize the CRB corresponding to the most-probable angle of the target, $\theta_{\mathrm{max}}=\mathrm{arg}\max p_{\Theta}(\theta)$, which is given by 
		$\mathrm{CRB}_{\theta}=\frac{1}{\frac{2N_{\mathrm{RF},\mathrm{R}}L\gamma}{N_{\mathrm{R}}\sigma_{\mathrm{S}}^2}\mathrm{tr}(\Dot{\bm{M}}^H(\theta_{\mathrm{max}})\bm{W}_{\mathrm{RF}}\bm{W}_{\mathrm{RF}}^H\Dot{\bm{M}}(\theta_{\mathrm{max}})
			\bm{V}_{\mathrm{RF}}\bm{R}_{\mathrm{BB}}\bm{V}_{\mathrm{RF}}^H)}$.		
		 $\bm{V}_{\mathrm{RF}}$, $\bm{R}_{\mathrm{BB}}$, and $\bm{W}_{\mathrm{RF}}^H$ are iteratively optimized under the proposed AO-based framework.\looseness=-1
	\end{itemize}
		\begin{figure}[t]
		\centering
		\includegraphics[width=0.46\textwidth]{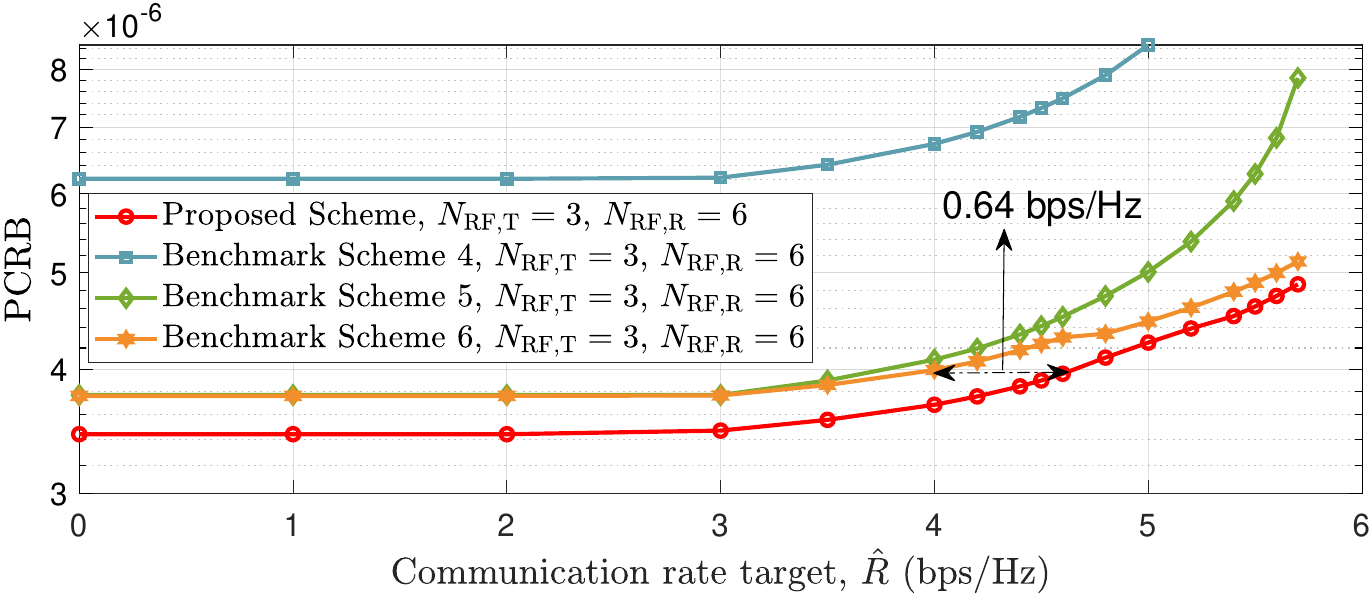}
		\vspace{-4mm}
		\caption{PCRB versus communication rate target for MIMO ISAC.}
		\label{R_VS_PCRB_ISAC_Benchmark}
	\end{figure}
	\begin{figure}[t]
		\centering
		\vspace{-4mm}
		\includegraphics[width=0.46\textwidth]{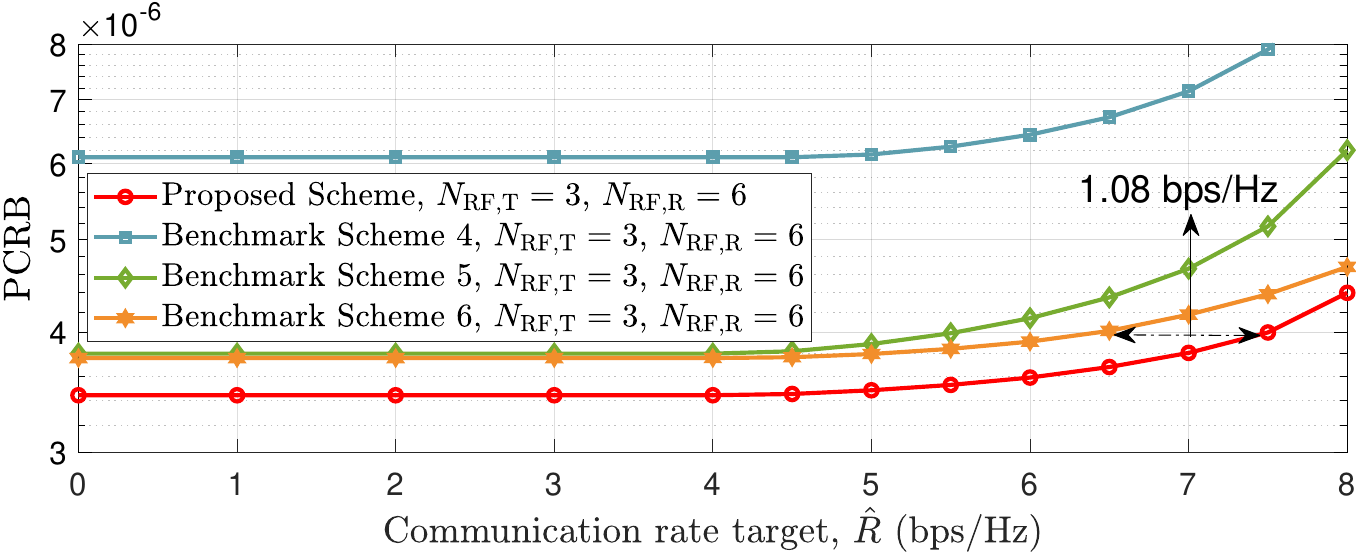}
		\vspace{-4mm}
		\caption{\hspace{-0.65ex}PCRB versus communication rate target for MIMO-OFDM ISAC.}
		\label{R_VS_PCRB_OFDM_Benchmark}
	\end{figure}
	Under a setup with $N_{\mathrm{RF},\mathrm{T}}=3$, $N_{\mathrm{RF},\mathrm{R}}=6$, and $N_{\mathrm{rand}}=100$, Figs. \ref{R_VS_PCRB_ISAC_Benchmark} and \ref{R_VS_PCRB_OFDM_Benchmark} show the PCRB versus rate target for MIMO ISAC and MIMO-OFDM ISAC, respectively. It is observed that the proposed scheme outperforms all the other benchmark schemes. Specifically, the random phase based scheme in Benchmark Scheme 4 has the most significant performance gap, which demonstrates the need of joint analog and digital beamforming design. Moreover, Benchmark Schemes 5 and 6 cannot achieve the performance of the proposed scheme due to the lack of tailored design based on the actual communication channel and the complete PDF of the target's angle. 
	\looseness=-1

\vspace*{-4mm}	
\section{Conclusions}\label{sec_conclusions}
\vspace*{-1mm}
This paper studied the hybrid beamforming optimization in a MIMO ISAC system, which aims to simultaneously communicate with a multi-antenna user and sense the unknown and random location information of a target exploiting its prior distribution information. 
Under hybrid analog-digital arrays at both the BS transmitter and the BS receiver, sensing is performed based on the analog-beamformed signals at the BS receiver, for which the performance was quantified by characterizing the PCRB. For the special case with only the sensing function, we derived the closed-form optimal solution to each entry in the analog beamforming matrices for PCRB minimization, and developed an AO-based algorithm which converges to a KKT point of the joint transmit/receive hybrid beamforming optimization problem. For narrowband MIMO ISAC systems, we studied the joint transmit/receive hybrid beamforming optimization problem to minimize the PCRB subject to a MIMO communication rate constraint. Despite its non-convexity, we devised an AO-based algorithm by leveraging WMMSE and FPP-SCA techniques. Then, we extended our PCRB characterization to wideband MIMO-OFDM ISAC systems and proposed a new AO-based algorithm that jointly optimized the individual transmit digital beamforming matrices over different sub-carriers and the common transmit/receive analog beamforming matrices for all sub-carriers.
The superiority of the proposed AO-based algorithms were validated via numerical results. It was also unveiled that the number of receive RF chains plays a more important role in the sensing performance compared to its transmit counterpart, and the optimal number of transmit RF chains under a given total number of transmit/receive RF chains increases with the communication rate target due to the PCRB-rate trade-off.

\bibliographystyle{IEEEtran}
\vspace{-3mm}
\bibliography{reference}

@inproceedings{bibGlobecom,
  author={Wang, Yizhuo and Zhang, Shuowen},
  booktitle={Proc. IEEE Global Commun. Conf. (Globecom)},
  title={Hybrid Beamforming Design for Integrated Sensing and Communication Exploiting Prior Information},
  year={2024},
  month ={Dec.},
  pages={4576-4581},
}

@INPROCEEDINGS{xu2023radar,
  author={Xu, Chan and Zhang, Shuowen},
  booktitle={Proc. IEEE Int. Symp. Inf. Theory (ISIT)}, 
  title={{MIMO} Radar Transmit Signal Optimization for Target Localization Exploiting Prior Information}, 
  year={Jun. 2023},
  volume={},
  number={},
  pages={310-315},
  doi={10.1109/ISIT54713.2023.10206741}
}

@article{xu2024mimo,
	title={{MIMO} integrated sensing and communication exploiting prior information},
	author={{C. Xu} and S. Zhang},
	journal={IEEE J. Sel. Areas Commun.},
	volume={42},
	number={9},
	pages={2306-2311},
	year={2024},
	month={Sep. },
	publisher={IEEE}
}

@INPROCEEDINGS{xu2024isac,
	author={Xu, Chan and Zhang, Shuowen},
	booktitle={Proc. IEEE Int. Symp. Inf. Theory (ISIT)}, 
	title={Integrated Sensing and Communication Exploiting Prior Information: {H}ow Many Sensing Beams are Needed?}, 
	year={Jul. 2024},
	pages={2802-2807}
}

@inproceedings{yao2024gc,
  author={Yao, Jiayi and Zhang, Shuowen},
  booktitle={Proc. IEEE Global Commun. Conf. (Globecom)},
  title={Optimal Transmit Signal Design for Multi-Target {MIMO} Sensing Exploiting Prior Information},
  year={2024},
  month ={Dec.},
  pages={4920-4925},
}

@article{hou2023optimal,
	title={Optimal Beamforming for Secure Integrated Sensing and Communication Exploiting Target Location Distribution}, 
	author={{K. Hou} and S. Zhang},
	journal={IEEE J. Sel. Areas Commun.},
	volume={42},
	number={11},
	pages={3125-3139},
        year={2024},
        month={Nov. },
        publisher={IEEE}
}

@Misc{yao2025optimal,
	title={Optimal Beamforming for Multi-Target Multi-User {ISAC} Exploiting Prior Information: {H}ow Many Sensing Beams Are Needed?}, 
	author={{J. Yao and S. Zhang}},
	year={[Online]. Available: https://arxiv.org/abs/2503.03560}
}

@inproceedings{hou2025base,
    author ={K. Hou and S. Zhang} ,
    title ={Base Station Placement Optimization for Networked Sensing Exploiting Target Location Distribution} ,
    booktitle = {Proc. IEEE Global Commun. Conf. (Globecom)},
      year={2025},
  month ={Dec.}
}

@article{zheng2025beyond,
  author={Zheng, Shuo and Zhang, Shuowen},
  journal={IEEE Trans. Cogn. Commun. Netw.}, 
  title={Beyond Diagonal Intelligent Reflecting Surface Aided Integrated Sensing and Communication}, 
  year={2025},
  volume={11},
  number={5},
  pages={2864-2878},
  month={Oct. }
}

@article{saad2020vision,
  title={A vision of 6{G} wireless systems: {A}pplications, trends, technologies, and open research problems},
  author={Saad, Walid and Bennis, Mehdi and Chen, Mingzhe},
  journal={IEEE Netw.},
  volume={34},
  number={3},
  pages={134--142},
  year={May. 2020},
  publisher={IEEE}
}

@article{liu2020joint,
	title={Joint transmit beamforming for multiuser {MIMO} communications and {MIMO} radar},
	author={Liu, Xiang and Huang, Tianyao and Shlezinger, Nir and Liu, Yimin and Zhou, Jie and Eldar, Yonina C},
	journal={IEEE Trans. Signal Process.},
	volume={68},
	pages={3929--3944},
	year={2020},
	month={Jun. },
	publisher={IEEE}
}

@article{hua2023optimal,
  title={Optimal transmit beamforming for integrated sensing and communication},
  author={Hua, Haocheng and Xu, Jie and Han, Tony Xiao},
  journal={IEEE Trans. Veh. Technol.},
  volume={72},
  number={8},
  pages={10588--10603},
  month={Aug. },
  year={2023},
  publisher={IEEE}
}

@article{liu2022cramer,
	title={Cram{\'e}r-{R}ao bound optimization for joint radar-communication beamforming},
	author={Liu, Fan and Liu, Ya-Feng and Li, Ang and Masouros, Christos and Eldar, Yonina C},
	journal={IEEE Trans. Signal Process.},
	volume={70},
	pages={240--253},
	year={2022},
	month={Dec. },
	publisher={IEEE}
}

@article{hua2024mimo,
	title={{MIMO} integrated sensing and communication: {CRB}-rate tradeoff},
	author={Hua, Haocheng and Han, Tony Xiao and Xu, Jie},
	journal={IEEE Trans. Wireless Commun.},
	year={2024},
	month={Apr. },
	volume={23},
	number={4},
	pages={2839--2854},
	publisher={IEEE},
}

@article{song2024cramer,
  title={Cram{\'e}r-rao bound minimization for {IRS}-enabled multiuser integrated sensing and communications},
  author={Song, Xianxin and Qin, Xiaoqi and Xu, Jie and Zhang, Rui},
  journal={IEEE Trans. Wireless Commun.},
  year={2024},
  month={Aug. },
  volume={23},
  number={8},
  pages={9714--9729},
  publisher={IEEE}
}

@article{tang2025dual,
  title={Dual-Function Beamforming Design For Multi-Target Localization and Reliable Communications},
  author={Tang, Bo and Li, Da and Wu, Wenjun and Saini, Astha and Babu, Prabhu and Stoica, Petre},
  journal={IEEE Trans. Signal Process.},
  year={2025},
  month={Jan. },
  volume={73},
  pages={559--573},
  publisher={IEEE}
}

@book{van1968detection,
	title={Detection, {E}stimation, and {M}odulation {T}heory, part {I}},
	author={Van Trees, Harry L},
	year={1968},
	publisher={Wiley, New York}
}

@Misc{attiah2024uplink,
	title={Uplink-Downlink Duality for Beamforming in Integrated Sensing and Communications}, 
	author={{K. M. Attiah and W. Yu}},
	year={[Online]. Available: https://arxiv.org/abs/2509.13661}
}

@article{liu2024ris,
  author={Liu, Yiming and Attiah, Kareem M. and Yu, Wei},
  journal={IEEE Trans. Wireless Commun.}, 
  title={{RIS}-Assisted Joint Sensing and Communications via Fractionally Constrained Fractional Programming},
  year={2025},
  month={Aug. },
  volume={25},
  pages={1674--1689}
}

@article{yu2016alternating,
  title={Alternating minimization algorithms for hybrid precoding in millimeter wave {MIMO} systems},
  author={Yu, Xianghao and Shen, Juei-Chin and Zhang, Jun and Letaief, Khaled B},
  journal={IEEE J. Sel. Topics Signal Process.},
  volume={10},
  number={3},
  pages={485--500},
  month={Apr. },
  year={2016},
  publisher={IEEE}
}

@article{el2014spatially,
  title={Spatially sparse precoding in millimeter wave {MIMO} systems},
  author={El Ayach, Omar and Rajagopal, Sridhar and Abu-Surra, Shadi and Pi, Zhouyue and Heath, Robert W},
  journal={IEEE Trans. Wireless Commun.},
  volume={13},
  number={3},
  pages={1499--1513},
  month={Mar. },
  year={2014},
  publisher={IEEE}
}

@article{sohrabi2016hybrid,
  title={Hybrid digital and analog beamforming design for large-scale antenna arrays},
  author={Sohrabi, Foad and Yu, Wei},
  journal={IEEE J. Sel. Topics Signal Process.},
  volume={10},
  number={3},
  pages={501-513},
  year={2016},
  month={Apr. },
  publisher={IEEE}
}

@article{sohrabi2017hybrid,
  title={Hybrid analog and digital beamforming for mm{W}ave {OFDM} large-scale antenna arrays},
  author={{F. Sohrabi} and {W. Yu}},
  journal={IEEE J. Sel. Areas Commun.},
  volume={35},
  number={7},
  pages={1432--1443},
  year={Jul. 2017},
  publisher={IEEE}
}

@inproceedings{liu2019hybrid,
	title={Hybrid beamforming with sub-arrayed {MIMO} radar: {E}nabling joint sensing and communication at mm{W}ave band},
	author={Liu, Fan and Masouros, Christos},
	booktitle={Proc. IEEE Int. Conf. Acoust., Speech Signal Process. (ICASSP)},
	pages={7770--7774},
	year={May. 2019},
}

@article{wang2022partially,
	title={Partially-connected hybrid beamforming design for integrated sensing and communication systems},
	author={Wang, Xinyi and Fei, Zesong and Zhang, J Andrew and Xu, Jie},
	journal={IEEE Trans. Commun.},
	volume={70},
	number={10},
	pages={6648--6660},
        month={Oct. },
	year={2022},
	publisher={IEEE}
}

@inproceedings{hu2022low,
  title={Low-{PAPR} {DFRC} {MIMO}-{OFDM} waveform design for integrated sensing and communications},
  author={Hu, Xiaoyan and Masouros, Christos and Liu, Fan and Nissel, Ronald},
  booktitle={Proc. IEEE Int. Conf. Commun. (ICC)},
  pages={1599--1604},
  year={May. 2022}
}

@article{wei2025precoding,
  title={Precoding optimization for {MIMO-OFDM} integrated sensing and communication systems},
  author={Wei, Zhiqing and Yao, Rubing and Yuan, Xin and Wu, Huici and Zhang, Qixun and Feng, Zhiyong},
  journal={IEEE Trans. Cogn. Commun. Netw.},
  volume={11},
  number={1},
  pages={288--299},
  month={Feb. },
  year={2025},
  publisher={IEEE}
}

@article{nguyen2024joint,
  title={Joint communications and sensing hybrid beamforming design via deep unfolding},
  author={Nguyen, Nhan Thanh and Nguyen, Ly V and Shlezinger, Nir and Eldar, Yonina C and Swindlehurst, A Lee and Juntti, Markku},
  journal={IEEE J. Sel. Topics Signal Process.},
  volume={18},
  number={5},
  pages={901--916},
  month={Jul. },
  year={2024},
  publisher={IEEE}
}

@article{liu2020joint1,
	title={Joint radar and communication design: {A}pplications, state-of-the-art, and the road ahead},
	author={Liu, Fan and Masouros, Christos and Petropulu, Athina P and Griffiths, Hugh and Hanzo, Lajos},
	journal={IEEE Trans. Commun.},
	volume={68},
	number={6},
	pages={3834--3862},
	month={Jun},
	year={2020},
	publisher={IEEE}
}

@article{shen2010fundamental,
  title={Fundamental limits of wideband localization—{P}art {I}: {A} general framework},
  author={Shen, Yuan and Win, Moe Z},
  journal={IEEE Trans. Inf. Theory},
  volume={56},
  number={10},
  pages={4956--4980},
  year={Oct. 2010},
  publisher={IEEE}
}

@book{kay1993fundamentals,
  title={Fundamentals of {S}tatistical {S}ignal {P}rocessing: {E}stimation {T}heory},
  author={Kay, Steven M},
  year={1993},
  publisher={Englewood Cliffs, NJ, USA: Prentice-hall}
}

@article{shi2011iteratively,
	title={An iteratively weighted {MMSE} approach to distributed sum-utility maximization for a {MIMO} interfering broadcast channel},
	author={Shi, Qingjiang and Razaviyayn, Meisam and Luo, Zhi-Quan and He, Chen},
	journal={IEEE Trans. Signal Process.},
	volume={59},
	number={9},
	pages={4331--4340},
	year={2011},
	month={Sep. },
	publisher={IEEE}
}

@article{mehanna2014feasible,
  title={Feasible point pursuit and successive approximation of non-convex {QCQP}s},
  author={Mehanna, Omar and Huang, Kejun and Gopalakrishnan, Balasubramanian and Konar, Aritra and Sidiropoulos, Nicholas D},
  journal={IEEE Signal Process. Lett.},
  volume={22},
  number={7},
  pages={804--808},
  year={2015},
  month={Jul. },
  publisher={IEEE}
}

@ARTICLE{zhang2005phase,
  author={Xinying Zhang and Molisch, A.F. and Sun Yuan Kung},
  journal={IEEE Trans. Signal Process.}, 
  title={Variable-phase-shift-based {RF}-baseband codesign for {MIMO} antenna selection}, 
  year={2005},
  month={Nov. },
  volume={53},
  number={11},
  pages={4091-4103},
}

@article{alkhateeb2014channel,
	title={Channel estimation and hybrid precoding for millimeter wave cellular systems},
	author={Alkhateeb, Ahmed and El Ayach, Omar and Leus, Geert and Heath, Robert W},
	journal={IEEE J. Sel. Topics Signal Process.},
	volume={8},
	number={5},
	pages={831--846},
	month={Oct. },
	year={2014},
	publisher={IEEE}
}

@book{CVX,
  title={Convex Optimization},
  author={Boyd, Stephen and Vandenberghe, Lieven},
  year={2004},
  publisher={Cambridge, U.K.: Cambridge Univ. Press}
}

@online{cvxtool,
  author="Grant, Michael and Boyd, Stephen",
  title="{(Jun. 2015). \textit{CVX: MATLAB Software for Disciplined Convex Programming}}",
  url="http://cvxr.com/cvx/",
}

@article{solodov1998convergence,
	title={On the convergence of constrained parallel variable distribution algorithms},
	author={Solodov, Michael V},
	journal={SIAM J. Optim.},
	volume={8},
	number={1},
	pages={187--196},
	month={Feb.},
	year={1998},
	publisher={SIAM}
}

@article{zhang2020capacity,
  title={Capacity characterization for intelligent reflecting surface aided {MIMO} communication},
  author={Zhang, Shuowen and Zhang, Rui},
  journal={IEEE J. Sel. Areas Commun.},
  volume={38},
  number={8},
  pages={1823--1838},
  month={Aug. },
  year={2020},
  publisher={IEEE}
}

@inproceedings{liu2025mimo,
  title={{MIMO} sensing beamforming design with low-resolution transceivers},
  author={Liu, Yiming and Yu, Wei},
  booktitle={Proc. IEEE Int. Conf. Commun. (ICC)},
  pages={608--613},
  month={Jun. },
  year={2025}
}
\vspace{-2mm}
\end{document}